\documentclass{pasj01_mod}

\Received{\today}
\Accepted{}
 
\usepackage{threeparttable,booktabs,color,amsmath}

\begin{document}

\title{Is the low-energy tail of shock-accelerated protons responsible for over-ionized plasma in supernova remnants?}

\author{Makoto \textsc{Sawada}\altaffilmark{1, 2}%
}
\altaffiltext{1}{Department of Physics, Rikkyo University, 3-34-1, Nishi-ikebukuro, Toshima-ku, Tokyo 171-8501, Japan}
\altaffiltext{2}{Center for Pioneering Research, RIKEN, 2-1 Hirosawa, Wako, Saitama 351-0198, Japan}
\email{makoto.sawada@rikkyo.ac.jp}

\author{Liyi \textsc{Gu},\altaffilmark{3, 4, 5, 6}}
\altaffiltext{3}{SRON Netherlands Institute for Space Research, Niels Bohrweg 4, 2333 CA Leiden, the Netherlands}
\altaffiltext{4}{Center for Pioneering Research, RIKEN, 2-1 Hirosawa, Wako, Saitama 351-0198, Japan}
\altaffiltext{5}{Leiden Observatory, Leiden University, PO Box 9513, 2300 RA Leiden, The Netherlands}
\altaffiltext{6}{Department of Physics, Tokyo University of Science, 1-3 Kagurazaka, Shinjuku-ku, Tokyo 162-8601, Japan}

\author{Ryo \textsc{Yamazaki}\altaffilmark{7, 8}}
\altaffiltext{7}{Department of Physical Sciences, Aoyama Gakuin University, 5-10-1 Fuchinobe, Sagamihara, Kanagawa 252-5258, Japan}
\altaffiltext{8}{Institute of Laser Engineering, Osaka University, 2-6 Yamadaoka,
Suita, Osaka 565-0871, Japan}


\maketitle

\begin{abstract}
Over-ionized, recombining plasma is an emerging class of X-ray bright supernova remnants (SNRs). This unique thermal state where the ionization temperature ($T_{\rm z}$) is significantly higher than the electron temperature ($T_{\rm e}$) is not expected from the standard evolution model assuming a point explosion in a uniform interstellar medium, requiring a new scenario for the dynamical and thermal evolution. A recently proposed idea attributes the over-ionization state to additional ionization contribution from the low-energy tail of shock-accelerated protons. However, this new scenario has been left untested, especially from the atomic physics point of view. We report calculation results of the proton impact ionization rates of heavy-element ions in ejecta of SNRs. We conservatively estimate the requirement for accelerated protons, and find that their relative number density to thermal electrons needs to be higher than $5~(T_{\rm e}/{\rm 1~keV})\%$ in order to explain the observed over-ionization degree at $T_{\rm z}/T_{\rm e} \ge 2$ for K-shell emission. We conclude that the proton ionization scenario is not feasible because such a high abundance of accelerated protons is prohibited by the injection fraction from thermal to non-thermal energies, which is expected to be $\sim 1\%$ at largest.
\end{abstract}


\section{Introduction}

Recent progresses in X-ray observations of supernova remnants (SNRs) have emphasized the importance of interaction between hot plasma of SNRs and the environments in understanding their dynamical and thermal evolution. One of the most articulate cases would be the discovery of over-ionized/recombining plasma from more than 20 Galactic middle-aged SNRs (e.g., \cite{Yamaguchi2009}, \cite{Sawada2012}). In the standard evolution model of SNRs where an SN is a point explosion in a uniform medium with a typical density of interstellar medium (ISM: $\sim$1~cm$^{-3}$), the dynamical evolution is described by the Sedov-Taylor solution \citep{Sedov1959}, while the thermal evolution is described as a combination of two relaxation processes: non-equipartition between electron and ion temperatures ($T_{\rm e}$ and $T_{\rm i}$, repectively) and non-equilibrium ionization (NEI: e.g., \cite{Masai1994}). Immediately after the shock passage, temperatures of ions and electrons are likely not equilibrated and are possibly mass proportional. Once $T_{\rm e}$ gets high enough through Coulomb heating with hot ions, electrons gradually ionize atoms from quasi-neutral initial state to the collisional ionization equilibrium (CIE) state determined solely by $T_{\rm e}$. Therefore, NEI in the standard model means under-ionized/ionizing plasma. The existence of over-ionized plasma strongly suggests that there is a process that is missing in the standard model but playing a major role in the actual thermal evolution of a particular class of SNRs.

Previous studies on the over-ionized SNRs have mainly focused on a rapid electron cooling as a mechanism responsible for producing over-ionized plasma. The cooling would be due either to a rapid adiabatic expansion (rarefaction) of ejecta after the blast-wave shock breaks out of dense circumstellar medium (CSM) \citep{Itoh1989, Shimizu2012} or a thermal conduction from hot plasma to interacting cold atomic and molecular clouds \citep{Kawasaki2002, Kawasaki2005}. Recent studies have often used spatial correlation as a key observational clue as to the physical origin of the over-ionized plasma. For instance, \citet{Yamaguchi2018} analyzed spatially resolved NuSTAR spectra of the SNR W\,49\,B and found that the electron temperature was positively correlated with the recombination timescale, an indicator of the electron density. This correlation supports the rarefaction scenario in the sense that the spatial difference of the electron temperature can be explained as the difference in the magnitude of rarefaction. On the other hand, \citet{Sano2021} used the same electron temperature distribution by NuSTAR but found another spatial correlation that rather supports the conduction cooling scenario. The lower electron temperature region was also associated with higher $^{12}$CO ($J=$2--1) molecular line emission measured by ALMA, indicating a higher impact of conduction cooling there. The origin of the over-ionized plasma in W\,49\,B is therefore not conclusively determined. The situation is more or less the same for other over-ionized SNRs.

More recently, an alternative scenario has been proposed. In the new scenario, the over-ionization is attained by extra-ionization of ions due to collisions with shock-accelerated protons \citep{Hirayama2019, Yamauchi2021} rather than by a rapid cooling of electrons. This is mainly motivated by the fact that most of the SNRs that harbor over-ionized plasma are bright GeV/TeV sources, indicating the existence of shock-accelerated protons. The low-energy continuation of the accelerated protons in the MeV--GeV range may have large cross sections for direct impact ionization of thermal ions and thus may alter the charge-state distribution (CSD) significantly. However, there have been no theoretical studies that attempted to reproduce the over-ionized plasma by incorporating the proton ionization, which leaves the scenario nearly untested. The only case so far is an examination from the energy budget point of view for a particular case of IC\,443 \citep{Okon2021}. Therefore, it is currently unknown even whether such an extra-ionization process is efficient enough to cause a significant change in the CSD. For all the over-ionized SNRs, the CSD of the element with strong K-shell RRC detection are dominated by He-like ions. Thus, contribution of accelerated protons to the K-shell ionization of these highly ionized atoms is of particular importance. 

In this paper, we present a theoretical test on the feasibility of attaining high CSD observed in over-ionized plasma in SNRs for the first time, using model calculations of the direct impact ionization of thermal ions with mildly energetic protons. The paper is organized as follows. In \S\ref{sec:model}, an overview of the model for the thermal evolution and the CSD calculation is given. In \S\ref{sec:xsec}, the methods for obtaining cross sections of proton impact ionization are given. Results of the calculations are presented in \S\ref{sec:result}. The feasibility of the proton ionization scenario and the origin of over-ionized/recombining plasma in SNRs are discussed in \S\ref{sec:disc}. Finally, the summary of this study is given in \S\ref{sec:summary}.

\section{Model overview} \label{sec:model}

\subsection{Thermal evolution}

In a hot plasma undergoing transient ionization, the CSD deviates from the one expected from the observed electron temperature ($T_{\rm e}$). A pseudo temperature called the ionization temperature ($T_{\rm z}$) is thus introduced to describe the CSD. A plasma with $T_{\rm z} < T_{\rm e}$ is under-ionized, while $T_{\rm z} > T_{\rm e}$ is over-ionized, in comparison to CIE. These transient ionization states are a consequence of an abrupt change either in the electron temperature or CSD, which will gradually approach to a CIE state through collisional relaxation. This process can be traced by using an NEI model. The simplest NEI calculation solves the time evolution of CSD assuming a constant electron temperature at $T_{\rm e}$. The degree of the relaxation is expressed by a single parameter, $\tau \equiv \int n_{\rm e} dt$, which is often called the ionization timescale for an ionizing plasma ($dT_{\rm z}/d\tau > 0$) or the recombination timescale for a recombining plasma ($dT_{\rm z}/d\tau < 0$).

\begin{figure}[htbp]
\begin{center}
\includegraphics[width=8cm]{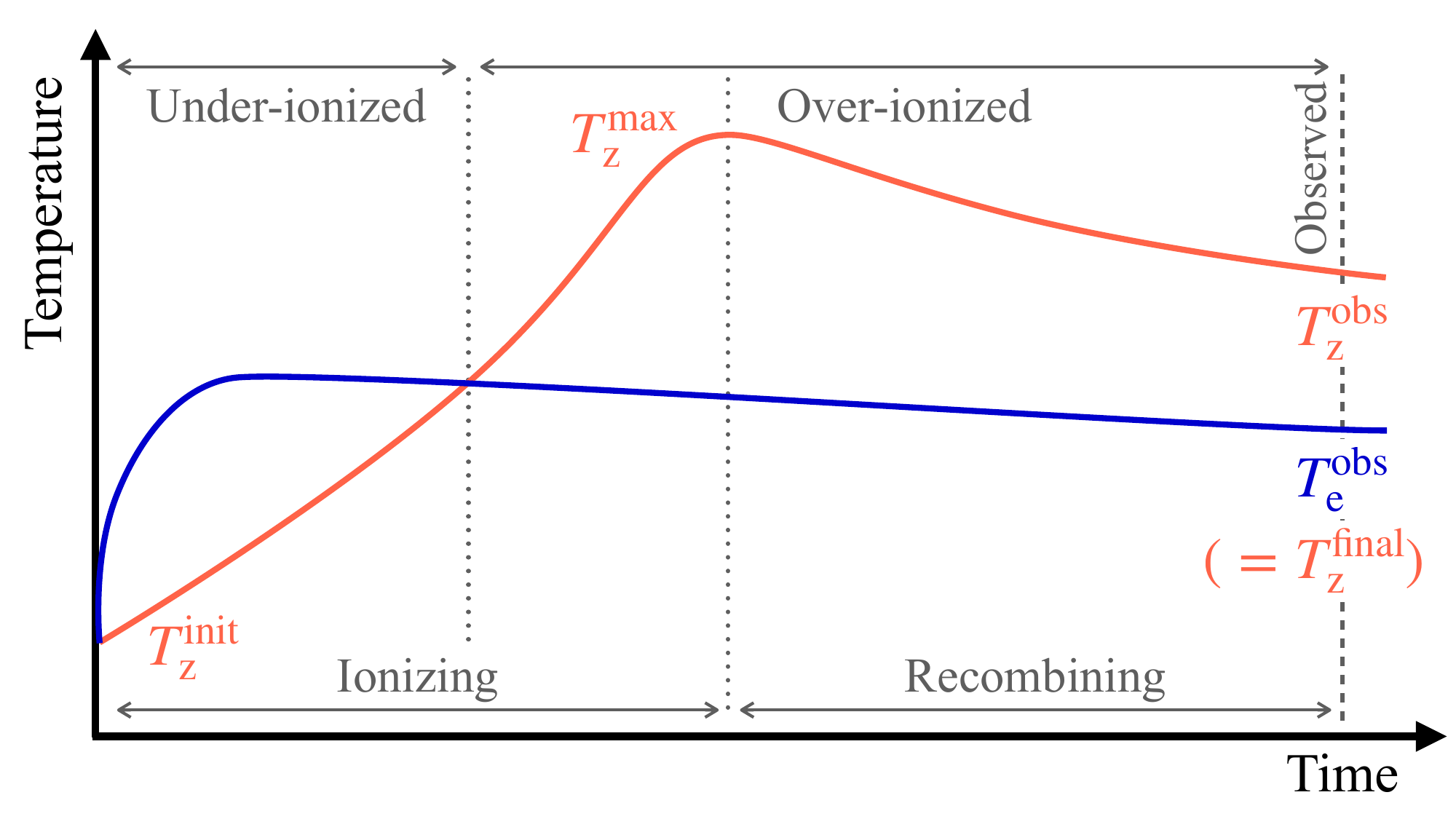}
\end{center}
\caption{A schematic of formation and evolution of a recombining plasma for the proton ionization scenario.}
\label{fig:schematic}
\end{figure}

The proton ionization scenario may consist of two evolutionary phases: the ionizing phase and recombining phase (Figure~\ref{fig:schematic}). In the ionizing phase, accelerated protons as well as thermal electrons contribute to the ionization of ions to raise $T_{\rm z}$. If the proton ionization rate is sufficiently high, $T_{\rm z}$ exceeds $T_{\rm e}$ at some point in this phase, resulting in an over-ionized state. The existence of the recombining phase is not necessarily required by the proton ionization scenario itself but suggested by observations; the overall spectra of recombining plasma in SNRs have been successfully reproduced with a recombining NEI model with non-zero recombination timescales in previous studies (e.g, \cite{Yamaguchi2012, Sawada2012}). This means that, the observed ionization temperature $T_{\rm z}^{\rm obs}$ has been somewhat reduced already from the maximum value $T_{\rm z}^{\rm max}$ attained in the ionizing phase. In this sense, observed deviations in the CSD from the CIE case, characterized by $T_{\rm z}^{\rm obs}$ and $T_{\rm e}^{\rm obs}$, should be regarded as a minimum change that the proton ionization scenario is required to explain. Unlike other scenarios, the proton ionization scenario does not involve any rapid cooling of electrons. Thus, as in the standard Sedov evolution, the gradual adiabatic cooling of electrons is expected to be nearly canceled by the Coulomb heating by thermal protons and ions \citep{Masai1994}, and the electron temperature is maintained to be almost constant throughout the thermal evolution. 

\begin{figure}[htbp]
\begin{center}
\includegraphics[width=8cm]{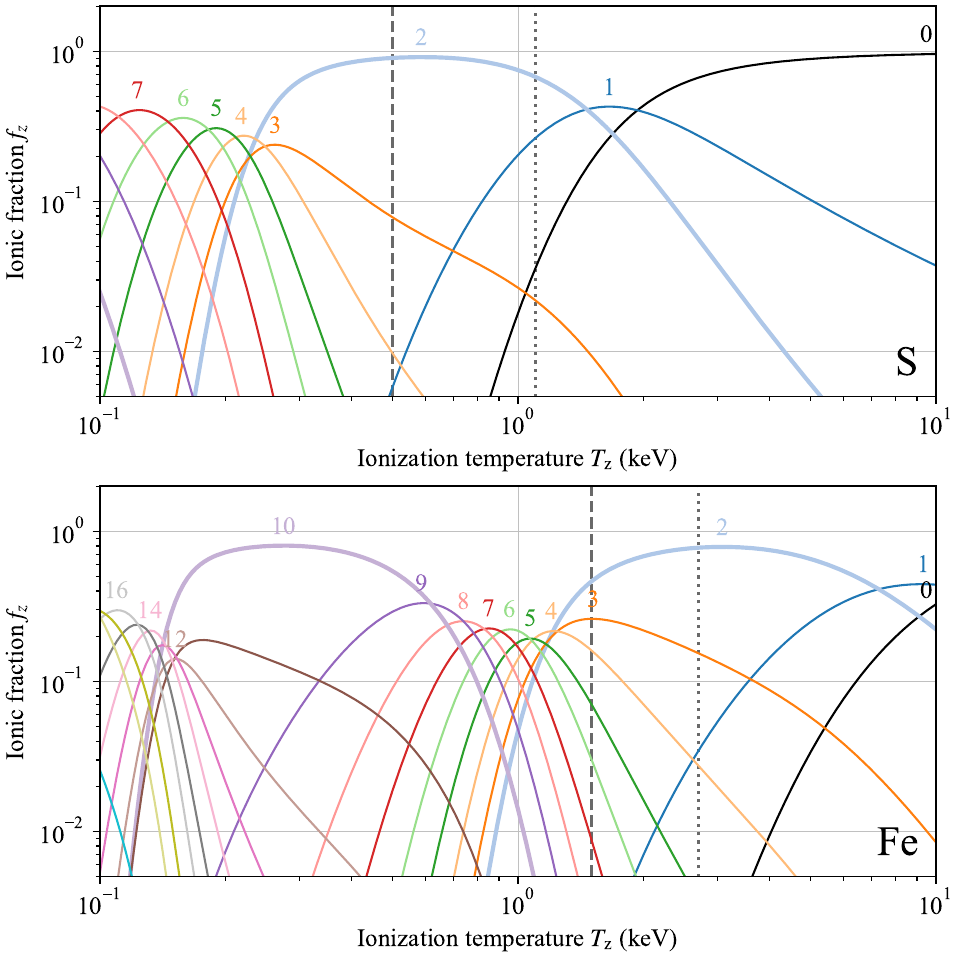}
\end{center}
\caption{The ionic fractions as a function on the ionization temperature for S (top) and Fe (bottom). The number attached to each curve shows the number of bound electrons (e.g., 2 for He-like and 10 for Ne-like). The dashed and dotted vertical lines indicate $T_{\rm e}^{\rm obs}$ and $T_{\rm z}^{\rm obs}$, respectively, whose values are of the average of IC\,443 and W\,28 for S and of W\,49\,B for Fe in Table~\ref{tab:obsts}.}
\label{fig:conrel}
\end{figure}

Given these conditions, the feasibility test we will carry out in this paper against the proton ionization scenario is to examine whether $T_{\rm z} \ge T_{\rm z}^{\rm obs}$ can be achieved with the additional proton ionization in otherwise-standard NEI calculations of the CSD assuming a constant electron temperature at $T_{\rm e}^{\rm obs}$. Table~\ref{tab:obsts} summarizes the observed thermal parameters for representative SNRs with over-ionized plasma. Note that, among the over-ionized SNRs, so far W\,49\,B has been the only case with $T_{\rm e}^{\rm obs} > 1$~keV and $T_{\rm z}^{\rm obs} > 2$~keV, and all others have similar parameters to IC\,443 or W\,28 with some variations. Therefore, our test will be made for the two cases, W\,49\,B-like one and IC\,443/W\,28-like one. The corresponding ionic fractions for the elements with strong K-shell RRC detection are shown in Figure~\ref{fig:conrel}. 

\begin{table}[htbp]
   \caption{Parameters for representative over-ionized SNRs.}
   \label{tab:obsts}
\begin{center}
   \begin{tabular}{lccll}
\toprule 
Object & $T_{\rm e}^{\rm obs}$ & $T_{\rm z}^{\rm obs}$ & Element & Ref.\footnotemark[$*$] \\
& (keV) & (keV) & with RRC \\
\midrule
W\,49\,B & 1.5 & 2.7 & Fe & 1 \\
IC\,443 & 0.6 & 1.2 & S & 2 \\
W\,28 & 0.4 & 1.0 & S & 3 \\
\bottomrule
\end{tabular}
\end{center}
\begin{tabnote}
\hangindent28pt\noindent
\hbox to20pt{\hss}\unskip%
   \footnotemark[$*$] 1: \citet{Ozawa2009}, 2: \citet{Yamaguchi2009}, \\and 3: \citet{Sawada2012}. 
\end{tabnote}
\end{table}

\subsection{Charge-state distribution}

We employ a simple model to calculate the CSD of hot plasma, where both electrons and protons are considered as contributors to collisional ionization. To investigate whether proton impact ionization alone can explain the over-ionized plasma in SNRs such as W49B and IC443, we do not consider the effects introduced in other scenarios, such as rapid electron cooling. Instead, we assume a standard, Sedov-like dynamical and thermal evolution, in which the spatial variation and temporal change of the electron temperature is not significantly large \citep{Masai1994}. 

The CSD can be obtained by solving a matrix differential equation (\cite{Masai1984}; \cite{Hughes1985}; \cite{Smith2010}):

\begin{equation}
\frac{d\boldsymbol{f}}{d\tau} = \textsf{A} \boldsymbol{f}, \label{eq:dfdtau}
\end{equation}

\noindent
where $\tau = \int n_{\rm e} dt$ is the ionization timescale and the column vector $\boldsymbol{f}$ is ionic fractions, i.e., 

\begin{equation}
\boldsymbol{f} = 
\begin{pmatrix}
f_0, & f_1, & \cdots, & f_j, & \cdots, & f_{Z-1}, & f_Z
\end{pmatrix}^T,
\end{equation}

\noindent
where $Z$ is the atomic number.
The matrix $\textsf{A}$ is the transition rate matrix, whose element $a_{ij}$ is given as follows

\begin{equation}
a_{ij} = 
\begin{cases}
-(\sum_i S_{ij} + \alpha_j)~&{\rm for}~i=j \\
S_{ij}~&{\rm for}~i>j \\
\alpha_j~&{\rm for}~i=j-1 \\
0 &{\rm otherwise}
\end{cases}~~~~,
\label{eq:aij}
\end{equation}

\noindent
where $S_{ij}$ is for the ionization that removed $(i-j)$ electrons, $\sum_i S_{ij}$ is for the total ionization, and $\alpha_j$ is for the recombination, all initiated from the ionization stage $j$. By deriving the eigenvectors $\boldsymbol{V_j}$ and eigenvalues $\lambda_j$ of the matrix $\textsf{A}$, the time evolution of the ionic fractions for each element is obtained as

\begin{equation}
\boldsymbol{f}(\tau) = \textsf{V} \left[ \exp(\lambda_j \tau) \right] \textsf{V}^{-1} \boldsymbol{f}_{\rm init},
\end{equation}

\noindent
where

\begin{equation}
\textsf{V} = 
\begin{pmatrix}
\boldsymbol{V_0}, \boldsymbol{V_1}, \cdots, \boldsymbol{V_j}, \cdots, \boldsymbol{V_{Z-1}}, \boldsymbol{V_Z}
\end{pmatrix}
\end{equation}

\noindent
is the matrix of eigenvectors, $\left[ \exp(\lambda_j \tau) \right]$ is the diagonal matrix, and $\boldsymbol{f}_{\rm init}$ is the initial ionic fractions. We calculate ionic fractions for $Z=1, \dots, 30$, with an initial condition where H and He are fully ionized, while the other elements are neutral. This choice of the initial condition does not affect calculation results in the the electron temperature range of our interest, where the dominant ionization stage is around He-like.

The ionization rate $S_{ij}$ has non-zero value even for $i > j+1$ because of multiple Auger ionization \citep{Kaastra1993}. It is obtained as

\begin{equation}
S_{ij} = \sum_k S_{k} \cdot \psi(k, i-j),
\end{equation}

\noindent
where $k=$ 1s, 2s, 2p$_{\frac{1}{2}}$, $\dots$, 4s is the subshell index, $S_{k}$ is the subshell ionization rate per electron density, $\psi(k, n)$ is the probability that $n - 1$ electrons are removed after a single ionization of the subshell $k$ (thus resulting in removing $n$ electrons in total). 

The difference of our model from the standard NEI is that the subshell ionization rate has the proton contribution as well as the electron contribution. If we write the coefficient and cross section for the partial ionization at the subshell $k$ due to proton direct impact as $S_{k}^{\rm p}$ and $\sigma_{{\rm p}k}$, respectively, 

\begin{equation}
S_{k}^{\rm p} = \frac{1}{n_{\rm e}} \int_{E_{\rm p}^{\rm min}}^{E_{\rm p}^{\rm max}} {\mathcal{N}}_{\rm p}(E_{\rm p})\,\sigma_{{\rm p}k}(E_{\rm p})\,v_{\rm p}\,dE_{\rm p} 
~~\propto~~\frac{n_{\rm p}^{\rm nt}}{n_{\rm e}} \label{eq:sion_p}
\end{equation}

\noindent
where $E_{\rm p}$, $v_{\rm p}$, and $\mathcal{N_{\rm p}}$, are the kinetic energy, speed, energy spectrum of the number density of the non-thermal protons, respectively, and 

\begin{equation}
n_{\rm p}^{\rm nt} = \int_{E_{\rm p}^{\rm min}}^{E_{\rm p}^{\rm max}} \mathcal{N}_{\rm p}\,dE_{\rm p}
\end{equation}

\noindent
is the integrated number density of the non-thermal protons over the energy.
The normalization by $n_{\rm e}$ in Eq.~\ref{eq:sion_p} is due to the parameterization $\tau = \int n_{\rm e} dt$ in Eq.~\ref{eq:dfdtau}.
For the proton energy distribution, we consider a power-law type, i.e., 

\begin{equation}
{\mathcal{N}}_{\rm p}(E_{\rm p}) \propto E_{\rm p}^{\Gamma}.
\end{equation}

\noindent
Since the relative number density $n_{\rm p}^{\rm nt}/n_{\rm e}$ and the power-law index $\Gamma$ are not known, we test several different values for these parameters. In general, both of these parameters can change in time, but we ignore such time dependencies of the proton energy distribution for simplicity. 

The remaining parameter is $\sigma_{{\rm p}k}$, the proton impact ionization cross section for each subshell. There are, however, no readily available data for proton impact ionization cross sections in a subshell-resolved form for various elements and ionization stages, which are required to calculate the time evolution of the CSD with the proton contribution. Therefore, we estimate these in the next section. 

For the electronic part of the ionization, consisting of direct impact ionization and excitation-autoionization, and for the recombination rate, consisting of radiative and dielectronic components, we use the same values as the SPEX \citep{Kaastra1996} version 3.08.00, which are based on the compilations by \citet{Urdampilleta2017} and \citet{Mao2016}, respectively. The scheme of solving the ionic fraction time evolution described above is also the same as SPEX. The only difference in our calculations from SPEX, other than the inclusion of the proton contribution to the ionization, is that we do not include contribution from charge exchange  \citep{Gu2016} to the ionization or recombination rates of our model. This is because it has significant contribution only at very low charge states. As is the case for the initial ionic fractions, any small difference in the early time evolution will not affect a later time evolution and does not make any difference in the CSD when the mean charge state reaches to the range of our interest, He-like or higher.

\section{Proton impact ionization} \label{sec:xsec}

\subsection{Approach}

In the classical theory of the inelastic collisions (e.g., \cite{Gryzinski1965a, Gryzinski1965b, Gryzinski1965c}), the ratio of velocity square between incident and field particles plays a major role in the characteristics of the collisions, thus, following parameters are introduced:

\begin{align}
u_{\rm p} &= \left(\frac{v_{\rm p}}{v_k}\right)^2 = \frac{E_{\rm p}}{ I_k} \cdot \frac{m_{\rm e}}{m_{\rm p}}, \\
u_{\rm e} &= \left(\frac{v_{\rm e}}{v_k}\right)^2 = \frac{E_{\rm e}}{I_k},
\end{align}

\noindent
where $m$, $E$ and $v$ are the masses, kinetic energies, and velocities, respectively, for protons and electrons with the subscripts p and e, respectively, $I_k$ is the ionization potential of the subshell $k$, and $v_k = \sqrt{2 I_k / m_{\rm e}}$ is the ``orbital'' velocity of a bound electron at the subshell $k$. 

Ionization cross sections for electrons and protons at equal velocities are expected to approach to each other at the high energy limit because the minimum momentum transfer becomes identical between these (e.g., \cite{Rudd1985}). By writing the proton and electron impact partial ionization cross sections for the subshell $k$ as $\sigma_{{\rm p}k}$ and $\sigma_{{\rm e}k}$, respectively, we expect

\begin{equation}
\sigma_{{\rm p}k}(u_{\rm p}) = \sigma_{{\rm e}k}(u_{\rm e})~~{\rm for}~~u_{\rm p} = u_{\rm e} \gg 1. \label{eq:sigmacal}
\end{equation}

To get reliable estimate for the subshell-resolved cross sections of the direct proton impact ionization for all ions of elements up to $Z=30$, we combine the model-based energy dependence of the cross section and the normalization of the cross section calibrated using the electron impact ionization cross section by invoking Eq.~\ref{eq:sigmacal}, subshell by subshell. This is essentially the same approach taken by \citet{Kocharov2000} and \citet{Kovaltsov2001}, in which the CSD of energetic Fe ions in solar corona were studied. 

As pointed out by \citet{Kovaltsov2001}, theoretical models of the proton direct impact ionization cross sections show different behaviors near the effective ionization threshold energy above which the cross section becomes significantly large. This leaves up to a-factor-of-two difference in their resultant ionization rates even after the normalization calibration with Eq~\ref{eq:sigmacal}. To see the systematic uncertainty due to the model ambiguity, we evaluate the cross sections using three different models as done by \citet{Kovaltsov2001}: the Bohr's cross section (\cite{Bohr1948, Knudsen1984, Kocharov2000}), the binary encounter model (BEM) cross section (\cite{Gryzinski1965c, McGuire1973, Peter1991}), and the Bates-Griffing relation (BGR) cross section \citep{Barghouty2000}. 

\subsection{Relativistic corrections}

In the velocity range where the ionization cross sections are matched between the proton and electron impacts, the Bethe term, $I_k^2 u \sigma_{k}(u) \propto \ln(u)$, dominates the cross section. In this range, it is important to include the relativistic effect. The relativistic correction factor $R$ for the Bethe term can be written as a function of two parameters, $\iota = I_k/(m_{\rm e} c^2)$ and $\epsilon = E/(mc^2)$ \citep{Gryzinski1965b, Gryzinski1965c, Quarles1976} as below

\begin{equation}
R(\epsilon) = \frac{2 + \iota}{2 + \epsilon} \left(\frac{1 + \epsilon}{1 + \iota}\right)^2 \left[ \frac{(\iota + \epsilon) (2 + \epsilon) (1 + \iota)}{\epsilon (2 + \epsilon) (1 + \iota)^2 + \iota (2 + \iota)} \right]^{\frac{3}{2}}, \label{eq:relcorfull}
\end{equation}

\noindent
where $m$ and $E$ are the mass and kinetic energy of incident particles. For all ions up to $Z=30$, $\iota \ll 1$ is satisfied for both the proton and electron impacts, and thus, Eq.~\ref{eq:relcorfull} is reduced to

\begin{equation}
R(\epsilon) \approx 2 \frac{(\epsilon + 1)^2}{\epsilon + 2}.
\end{equation}

\noindent
With $\epsilon = u \cdot \iota$, for the mildly relativistic energy of $\epsilon = E/(m c^2) \lesssim 1$, the correction factor can be approximated by

\begin{equation}
R(u) \approx 1 + \frac{3}{2} (u \cdot \iota) + \frac{1}{4} (u \cdot \iota)^2. \label{eq:relcor}
\end{equation}

\noindent
This form can be used commonly for protons as $R_{\rm p}(u_{\rm p})$ or electrons as $R_{\rm e}(u_{\rm e})$.

\subsection{Electron impact ionization cross section}

For the electron impact ionization cross sections to calibrate the proton impact ionization cross sections, \citet{Kocharov2000} and \citet{Kovaltsov2001} used the expression by \citet{Arnaud1985},

\begin{align}
I_k^2 u_{\rm e} \sigma_{{\rm e}k}(u_{\rm e}) = {} & A_k \left(1 - \frac{1}{u_{\rm e}}\right) + B_k \left(1 - \frac{1}{u_{\rm e}}\right)^2 \nonumber \\
& + C_k \ln(u_{\rm e}) + E_k \frac{\ln(u_{\rm e})}{u_{\rm e}}, \label{eq:a92}
\end{align}

\noindent
where the coefficients $A_k$, $B_k$, $C_k$, and $E_k$ are given by \citet{Arnaud1985} for H, He, C, N, O, Ne, Na, Mg, Al, Si, S, Ar, Ca, Fe, and Ni, while those for Fe have been update by \citet{Arnaud1992}.
For our study, the updated compilation by \citet{Urdampilleta2017} is used in favor of both the advanced experimental data included and the consistency with the cross sections used for the electronic part of the ionization. \citet{Urdampilleta2017} used a modified expression to fit the experimental data,

\begin{align}
I_k^2 u_{\rm e} \sigma_{{\rm e}k}(u_{\rm e}) = {} & A_k \left(1 - \frac{1}{u_{\rm e}}\right) + B_k \left(1 - \frac{1}{u_{\rm e}}\right)^2 \nonumber \\
& + C_k R_{\rm e}(u_{\rm e}) \ln(u_{\rm e}) \nonumber \\
& + D_k \frac{\ln(u_{\rm e})}{\sqrt{u_{\rm e}}} + E_k \frac{\ln(u_{\rm e})}{u_{\rm e}}, \label{eq:u17}
\end{align}

\noindent
where the updated coefficients $A_k$, $B_k$, $C_k$, $D_k$, and $E_k$ are derived for all elements of $Z=1$--$30$.
There are two major differences from the formalization by \citet{Arnaud1985}. The first is the addition of a new term proportional to $\ln({u_{\rm e}}) / \sqrt{u_{\rm e}}$, which is required to keep the coefficient $C_k$ identical to the Bethe value, while allowing the model to reproduce experimental cross sections in a wide range of the electron energy. The second is the introduction of the relativistic correction $R_{\rm e}$ for the Bethe term (Eq.~\ref{eq:relcor}). 

\subsection{Proton impact ionization cross sections}

\subsubsection{The Bohr cross section}

For highly ionized ions with $I_k \ge 16 I_0$, where $I_0 = 13.6$~eV is the hydrogen ionization potential, the Bohr's cross section can be expressed as follows \citep{Kocharov2000}

\begin{equation}
I_k^2 u_{\rm p} \sigma_{{\rm p}k}^{\rm Bohr} = A_0 n_k \sigma_0 \left[ 1 - \frac{1}{4u_{\rm p}} + \delta \ln(4u_{\rm p}) \right], \label{eq:bohr}
\end{equation}

\noindent
where $\sigma_0 = 4\pi a_0^2I_0^2 = 6.56 \times 10^{-14}$~cm$^2$~eV$^2$, $a_0 = 5.292 \times 10^{-9}$~cm is the Bohr radius, and $n_k$ is the number of electrons in the subshell $k$. From the equivalence with the electron cross section (Eq.~\ref{eq:a92}) at the high-energy limit (Eq.~\ref{eq:sigmacal}), the normalization factors $A_0$ and $\delta$ are determined to be:

\begin{align}
A_0 &= \frac{A_k + B_k - C_k \ln(4)}{n_k \sigma_0}, \label{eq:bohr_a0} \\
\delta &= \frac{C_k}{A_k + B_k - C_k \ln(4)}. \label{eq:bohr_delta}
\end{align}

To make the Bohr cross section compatible with the updated electron cross section by \citet{Urdampilleta2017}, we modify the expression as follows:

\begin{align}
I_k^2 u_{\rm p} \sigma_{{\rm p}k}^{\rm Bohr} = A_0 n_k \sigma_0 \left[ 1 - \frac{1}{4u_{\rm p}} + \delta \ln(4u_{\rm p}^{R_{\rm p}}) \right],
\end{align}

\noindent
where $R_{\rm p}$ is the relativistic correction factor (Eq.~\ref{eq:relcor}) for protons, which assures $I_k^2 u_{\rm p} \sigma_{{\rm p}k}^{\rm Bohr} \approx C_k R_{\rm p} \ln(u_{\rm p})$ for $u_{\rm p} \gg 1$. 

\subsubsection{The BEM cross section}

The BEM cross section is given by

\begin{equation}
I_k^2 u_{\rm p} \sigma_{{\rm p}k}^{\rm BEM} = n_k \sigma_0 F(u_{\rm p}). \label{eq:bem}
\end{equation}

\noindent
For a large $u_{\rm p}$ satisfying $u_{\rm p} > u_{\rm thre}$,

\begin{equation}
F(u_{\rm p}) = \alpha^{3/2} (1 - \beta) (1 - \beta^{1+u_{\rm p}}) \left[ \alpha + 2\delta(1+\beta) \ln(e + \sqrt{u_{\rm p}})\right], \label{eq:bemforg}
\end{equation}

\noindent
where 

\begin{align}
\alpha &= \frac{u_{\rm p}}{1 + u_{\rm p}}, \\
\beta &= \frac{1}{4 (u_{\rm p} + \sqrt{u_{\rm p}})},
\end{align}

\noindent
and

\begin{equation}
u_{\rm thre} = \frac{3 - 2\sqrt{2}}{4}
\end{equation}

\noindent
is the threshold value of $u_{\rm p}$, which corresponds to the minimum proton energy required to ionize the bound electron with the mean velocity in the orbit at the ionization potential $I_k$. 
Below this energy ($u_{\rm p} < u_{\rm thre}$), the ionization may still take place due to the continuous velocity distribution of the bound electron, with the cross section characterized by \citet{Gryzinski1965b} and \citet{Gryzinski1965c} as

\begin{equation}
F(u) \approx \frac{4}{15} u_{\rm p}^3
\end{equation}

\citet{Kovaltsov2001} associated $\delta$ to the electron cross section coefficients as $\delta = C_k/(\sigma_0 n_k)$, which implies that the Bethe term ($\sim \ln{u}$) in Eq.~\ref{eq:bemforg} is the only non-negligible component at a high-energy range ($u \gg 1$). In other words, if we rewrite Eq.~\ref{eq:bemforg} as 
\begin{equation}
F(u_{\rm p}) = F_1(u_{\rm p}) + F_2(u_{\rm p}) + F_3(u_{\rm p}), 
\end{equation}

\noindent
where
\begin{align}
F_1(u_{\rm p}) &= \alpha^{3/2} (1 - \beta) (1 - \beta^{1+u_{\rm p}}) \cdot \alpha, \\
F_2(u_{\rm p}) &= \alpha^{3/2} (1 - \beta) (1 - \beta^{1+u_{\rm p}}) \cdot 2\delta \ln(e + \sqrt{u_{\rm p}}), \\
F_3(u_{\rm p}) &= \alpha^{3/2} (1 - \beta) (1 - \beta^{1+u_{\rm p}}) \cdot 2\delta \ln(e + \sqrt{u_{\rm p}}) \cdot \beta, 
\end{align}

\noindent
it was assumed that $F_1(u_{\rm p}) \sim F_3(u_{\rm p}) \sim 0$ at $u \gg 1$. However, for a fiducial value of $\delta = 1/3$ and $u$ ranging from $10$ to $10^5$, we find that this is not the case. While $F_3$ is $\approx 0$, $F_1$ is nearly constant at $\approx 1$, which is not negligible in comparison with the dominant component $F_2$ ranging from $\sim 1$--$4$. Thus, in associating the BEM cross section parameters to the electron cross section coefficients, we modify the expression to have the overall normalization $A_0$ analogous to the Bohr cross section:

\begin{equation}
I_k^2 u_{\rm p} \sigma_{{\rm p}k}^{\rm BEM} = A_0 n_k \sigma_0 F(u_{\rm p}), \label{eq:bemmod}
\end{equation}

\noindent
which yields

\begin{align}
A_0 &= \frac{A_k + B_k}{n_k \sigma_0}, \label{eq:bem_a0} \\ 
\delta &= \frac{C_k}{A_k + B_k}. \label{eq:bem_delta}
\end{align}

\noindent
As was the case for the Bohr cross section, we also modify Eq.~\ref{eq:bemforg} to have the relativistic correction factor for the compatibility with the electron cross section by \citet{Urdampilleta2017}:
\begin{align}
F(u_{\rm p}) = {} & \alpha^{3/2} (1 - \beta) (1 - \beta^{1+u_{\rm p}}) \nonumber \\
& \times \left[ \alpha + 2 R_{\rm p} \delta(1+\beta) \ln(e + \sqrt{u_{\rm p}})\right], \label{eq:bemfmod}
\end{align}

\noindent
which assures $I_k^2 u_{\rm p} \sigma_{{\rm p}k}^{\rm BEM} \approx C_k R_{\rm p} \ln(u_{\rm p})$ for $u_{\rm p} \gg 1$.

\subsubsection{The BGR cross section}

The BGR cross section is a simple estimation of proton-impact ionization cross section from the electron-impact cross section using a direct relation between the cross sections as follows \citep{Barghouty2000}:

\begin{equation}
\sigma_{{\rm p}k}^{\rm BGR}(u_{\rm p}) = \frac{(1 + 4u_{\rm p})^3}{4u_{\rm p}(16u_{\rm p}^2 + 4u_{\rm p} - 2)} \sigma_{{\rm e}k}(u_{\rm e}^{\prime}), \label{eq:bgr}
\end{equation}

\noindent
which satisfies $\sigma_{{\rm p}k}^{\rm BGR}(u_{\rm p}) \approx \sigma_{{\rm e}k}(u_{\rm e}^{\prime})$ for $u_{\rm p} \gg 1$,

\noindent
where
\begin{equation}
u_{\rm e}^{\prime} = \left(1 + \frac{1}{4u_{\rm p}} \right)^2 u_{\rm p}.
\end{equation}

\noindent
In this case, the conversion naturally includes the relativistic correction when the electron impact cross section of \citet{Urdampilleta2017} is used. One can find that, for $u_{\rm p} \gg 1$, $u_{\rm e}^{\prime} \approx u_{\rm p}$ and $I_k^2 u_{\rm p} \sigma_{{\rm p}k}^{\rm BGR} \approx C_k R_{\rm p} \ln(u_{\rm p})$.


\section{Results} \label{sec:result}

\subsection{Ionization cross sections}

\begin{figure*}[htbp]
\begin{center}
\includegraphics[width=16cm]{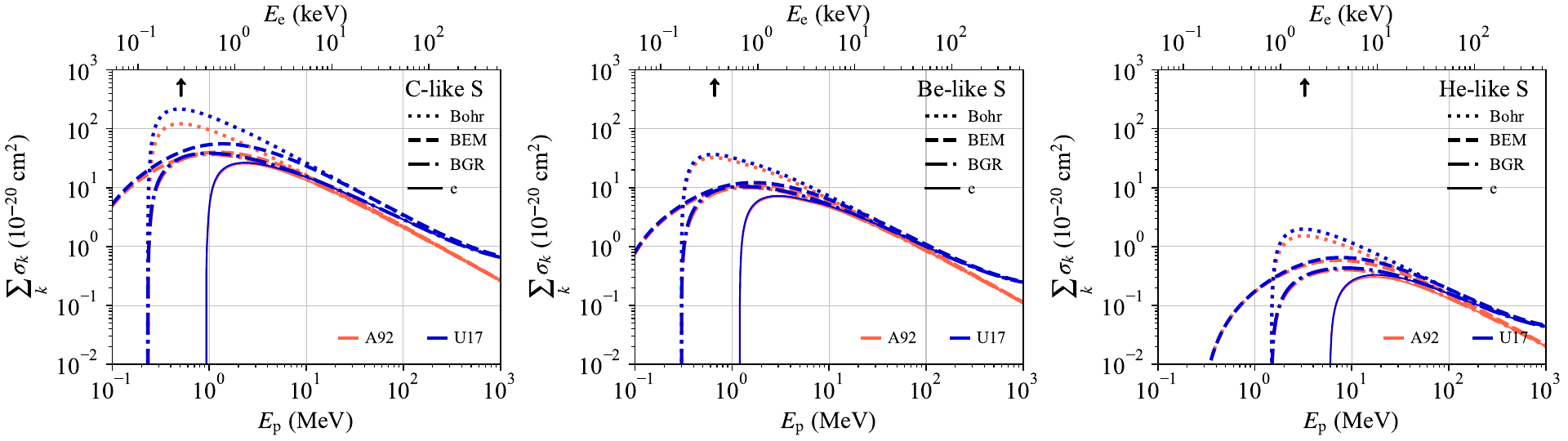}
\includegraphics[width=16cm]{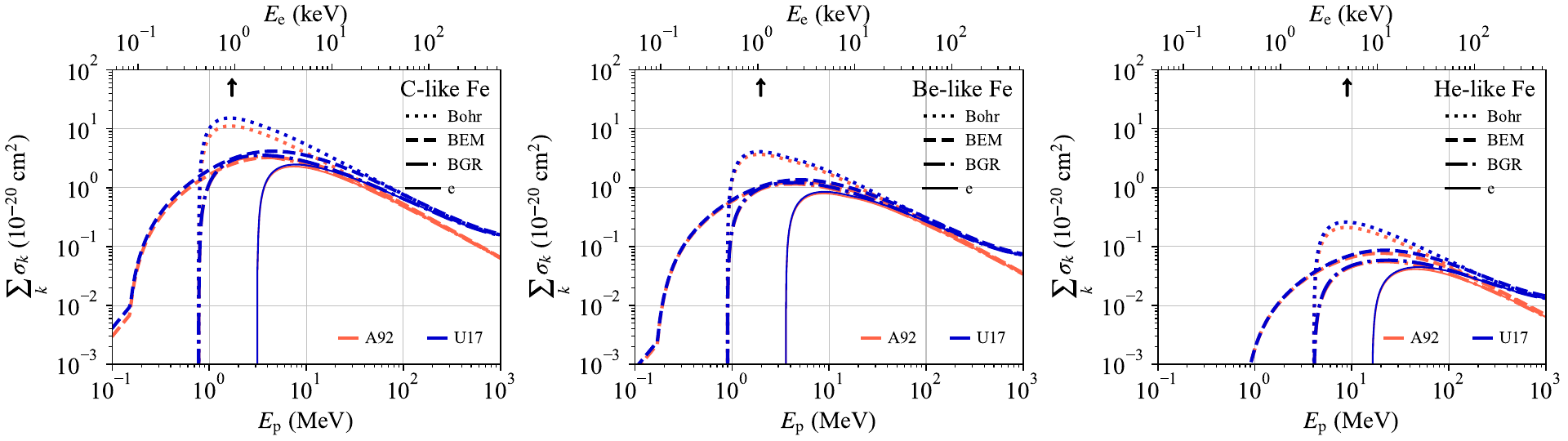}
\end{center}
\caption{Total ionization cross sections for C-like (left), Be-like (middle), and He-like (right) ions of S (top) and Fe (bottom). The three models of proton impact cross sections are shown with thick lines, while the electron impact cross sections are shown with thin solid lines. Colors correspond to different atomic data, A92 in red for \citet{Arnaud1992} and U17 in blue for \citet{Urdampilleta2017}. The black arrows show the ionization potential $I_k$ corresponding to $u_{\rm e} = 1$ for electrons (top axes) and $u_{\rm p} = 1$ for protons (bottom axes) for the outermost subshell.}
\label{fig:xsec}
\end{figure*}

Figure~\ref{fig:xsec} shows the results of the total proton impact ionization cross sections for selected ions of S and Fe as a function of the proton energy $E_{\rm p}$. The reference electron impact cross sections are also shown as a function of the electron energy $E_{\rm e}$, whose range is shifted such that $u_{\rm p} = u_{\rm e}$ at the same position in the horizontal direction in the figure. The three models of the proton impact ionization show significant variation near the effective threshold energy. There is a systematic trend of the Bohr cross section being the largest and the BGR being the smallest due to the difference near the ionization potential (the black arrow), and the BEM being in the middle due to the similar cross section above the ionization potential but with the lower threshold and thus wider range compared to the BGR model. On the other hand, these converge on each other and with the electron cross section at high energy limit of $u_{\rm p} \gg 1$ by design. It is commonly seen that the proton impact cross sections using \citet{Urdampilleta2017} as the reference electron impact cross sections are higher at the high-energy end due to the relativistic correction and are similar or slightly higher in lower energies compared to those based on \citet{Arnaud1992}. Hereafter, we will concentrate on the results using \citet{Urdampilleta2017} as it is one of the state-of-the-art compilation of the electron impact ionization cross sections. 

We note that, in the calculations of the Bohr and BEM cross sections, there are some cases where the normalization parameter $A_0$ became negative. The negative normalizations are found for outer electrons at or above the 2p subshells, and are limited to low-ionization ions of F-like or lower. Most cases are associated with the cross section parameters with large values of $D_k$ in Eq.~\ref{eq:u17}. In such a case, the term with the coefficient $D_k$ may not be negligible even at high energies, leading to a large error in the asymptotic relations in Eqs.~\ref{eq:bohr_a0}--\ref{eq:bohr_delta} and \ref{eq:bem_a0}--\ref{eq:bem_delta} and therefore invalid $A_0$ values. We treated the subshell cross sections with negative $A_0$ as 0 in the following calculations of ionization rates. Because the dominant ionization states are He-like for the observed ionization temperature (Figure~\ref{fig:conrel}), the underestimation of partial cross sections for low-ionization ions only increases the required ionization timescale $\tau$ to achieve a given CSD, and therefore this does not affect the maximum charge state achievable with the proton ionization at the large $n_{\rm e} t$ limit, which we will discuss later. 

\subsection{Ionization rates} \label{sec:result_rates}

Ionization rates are slightly affected by the choice of the energy range over which the proton energy distribution is integrated (between $E_{\rm p}^{\rm min}$ and $E_{\rm p}^{\rm max}$ in Eq.~\ref{eq:sion_p}). Because the ionization enhancement for ions such as He-like S and He-like Fe is most crucial to realize the observed over-ionization state, the minimum energy $E_{\rm p}^{\rm min}$ is taken at 1~MeV, which is just below the energy at which the ionization cross section becomes significant for He-like S with the Bohr and BGR models (Figure~\ref{fig:xsec}). The maximum energy $E_{\rm p}^{\rm max}$ is set to 1~GeV, to which the mildly relativistic approximation (Eq.~\ref{eq:relcor}) is valid. 

\begin{figure}[htbp]
\begin{center}
\includegraphics[width=8cm]{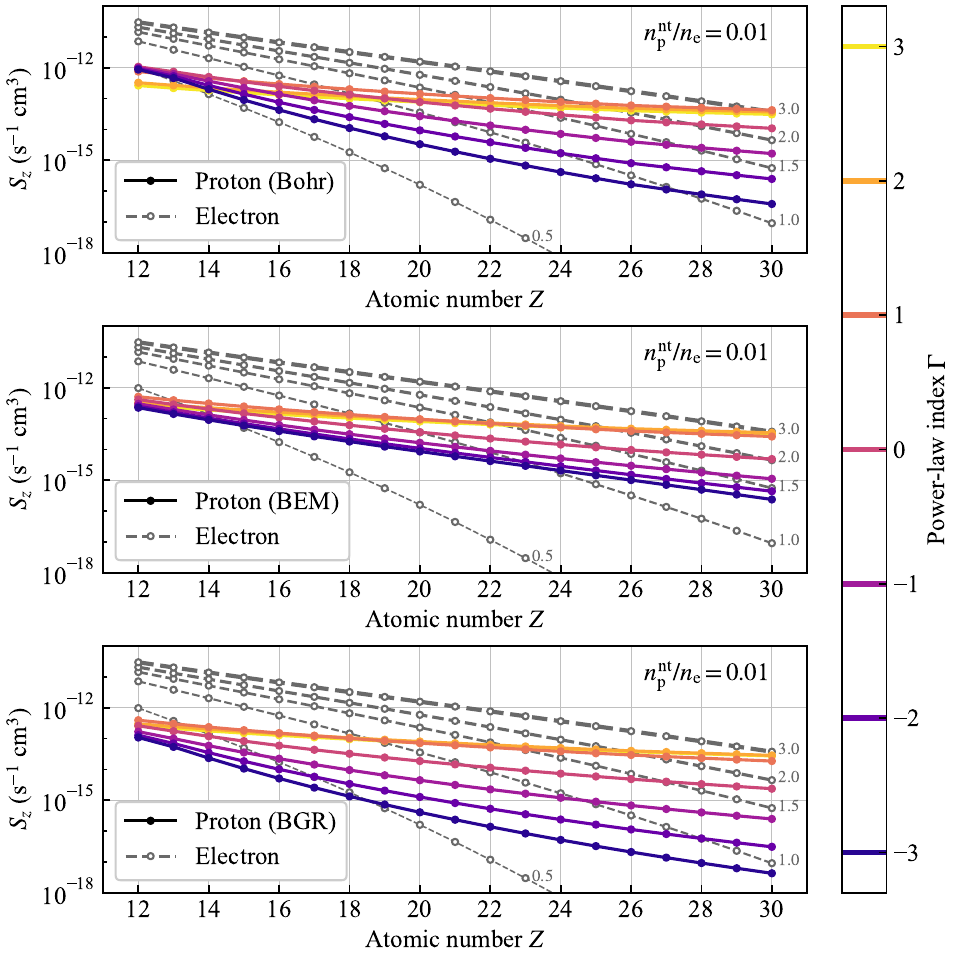}
\end{center}
\caption{Ionization rates for He-like ions of various elements for proton ionization (filled circles with colored solid curves) and for electron ionization (open circles with gray dashed curves). The three panels are for different proton ionization cross section models, Bohr (top), BEM (middle), and BGR (bottom), all assuming the relative density of $n_{\rm p}^{\rm nt} / n_{\rm e} = 0.01$. The colors correspond to different power-law index $\Gamma$ for the proton energy distribution. The curves for the electron ionization are for $kT_{\rm e} =$ 0.5, 1.0, 1.5, 2.0, and 3.0~keV from bottom to top.}
\label{fig:rate_elemdep}
\end{figure}

Figure~\ref{fig:rate_elemdep} shows the elemental dependence of the ionization rates for He-like ions with the relative density of $n_{\rm p}^{\rm nt}/n_{\rm e} = 0.01$ and various values of the power-law index $\Gamma$. The $\Gamma$-dependence of the proton ionization rate is milder compared to the $kT_{\rm e}$-dependence of the electron ionization rate. This is mainly attributable to the difference in the range of the particle energy above the (effective) ionization threshold; a power-law distribution can populate more ionizing particles above the energy at which the ionization cross section is significantly large, even for heavy elements like Fe. Similarly, the difference in the elemental dependence between the proton cross sections can be explained; the Bohr and BGR cross sections having higher effective ionization thresholds show steeper elemental dependences than the BEM cross section. For any elements considered here ($Z \ge 12$), the proton ionization rate is maximized when $\Gamma \approx +1.0$.

\begin{figure}[htbp]
\begin{center}
\includegraphics[width=8cm]{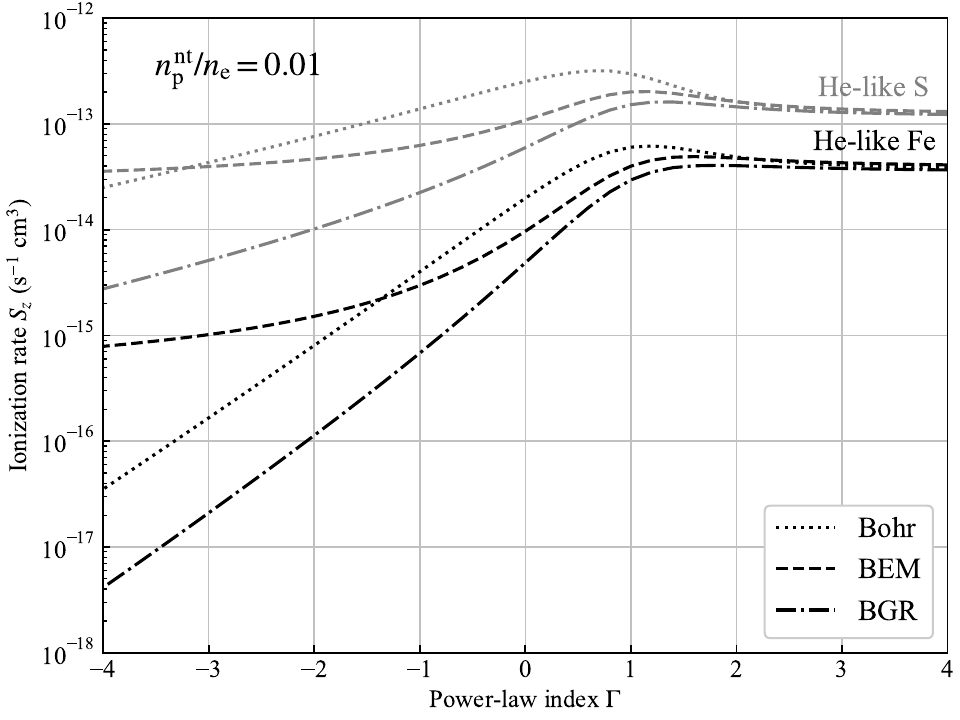}
\end{center}
\caption{The proton ionization rate dependence on the power-law index $\Gamma$ of the proton energy distribution for He-like S (gray) and He-like Fe (black).}
\label{fig:rate_einddep}
\end{figure}

Figure~\ref{fig:rate_einddep} shows the dependence on the power-law index $\Gamma$ for He-like S and He-like Fe. As seen in Figure~\ref{fig:rate_elemdep}, for the both cases, the ionization rates are maximized at $\Gamma \approx +1$. The three models show a large variation with a large negative $\Gamma$ reflecting the difference in the structure near the effective ionization threshold, while these converge well with a large positive index due to the common condition to all models (Eq.~\ref{eq:sigmacal}).

\subsection{Ionic fractions} \label{sec:ionfrac}

\begin{figure*}[htbp]
\begin{center}
\includegraphics[width=8cm]{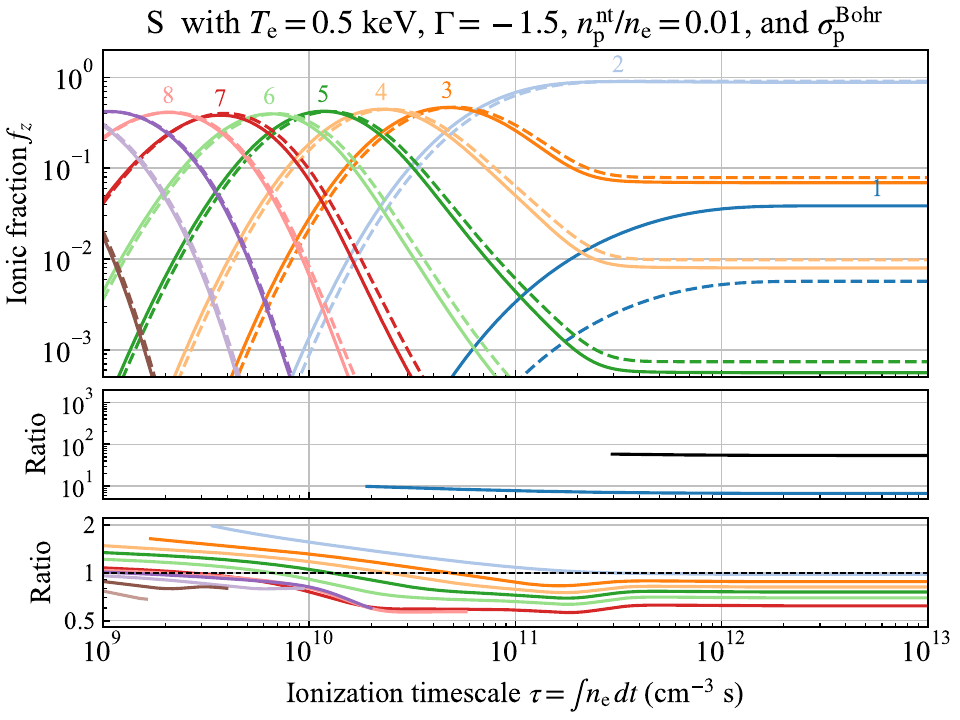}
\includegraphics[width=8cm]{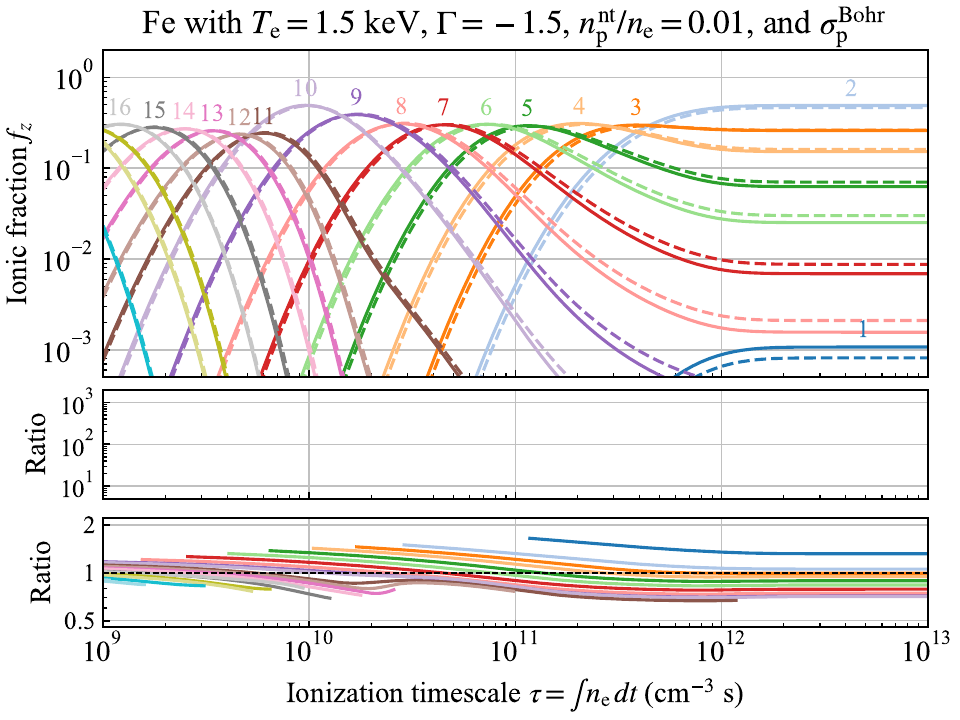}
\includegraphics[width=8cm]{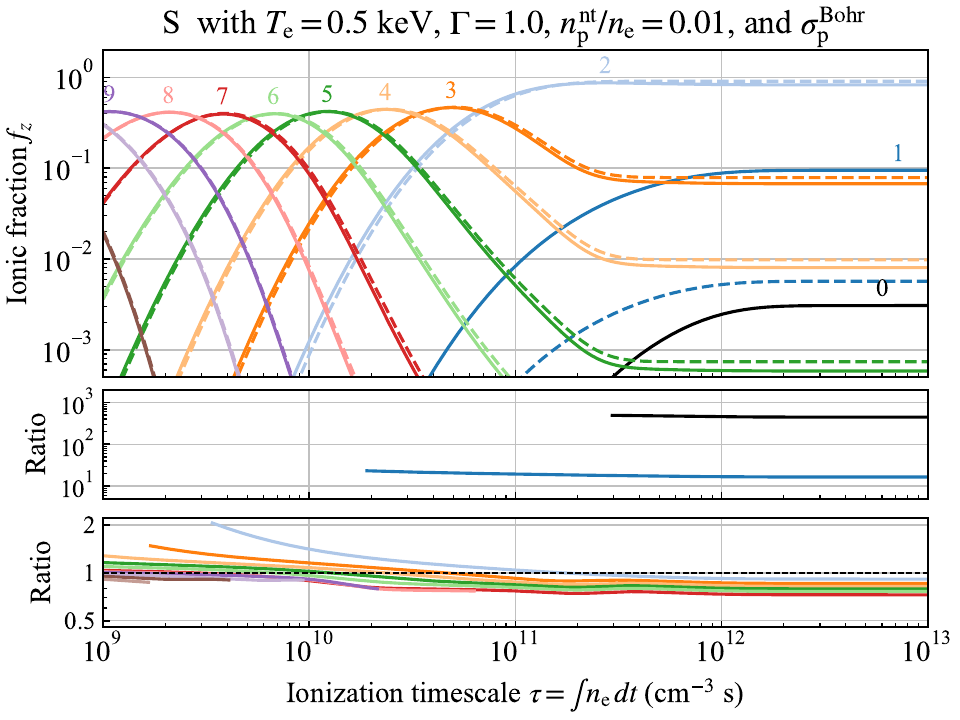}
\includegraphics[width=8cm]{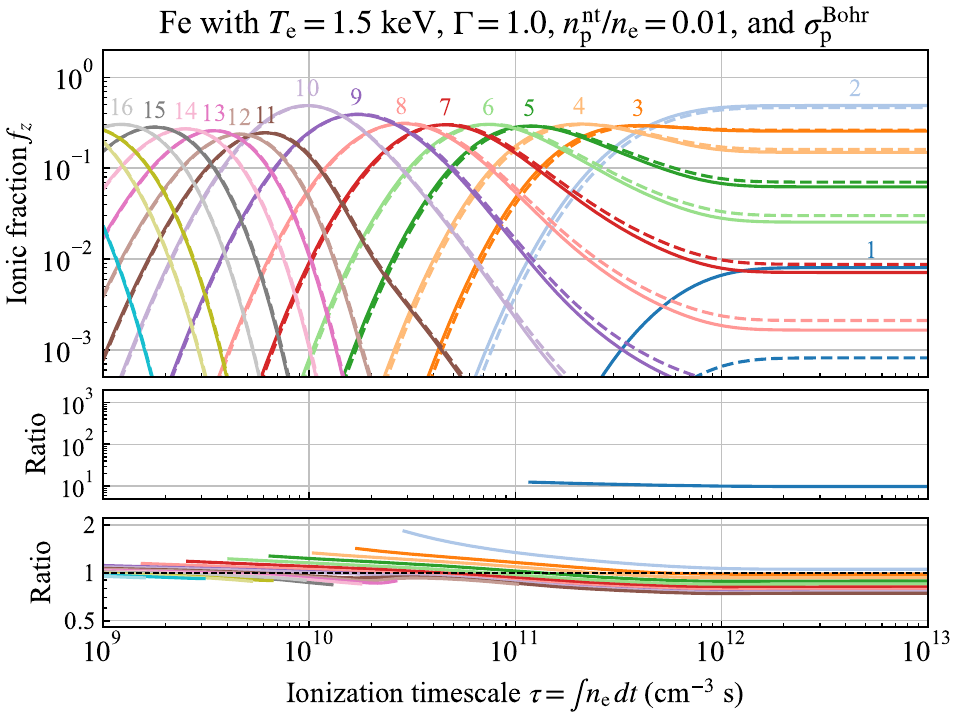}
\end{center}
\caption{The time evolution of ionic fractions with the proton impact ionization (solid) compared with thermal electron ionization only (dashed). The left subfigures are for S with $T_{\rm e} = 0.5$~keV, while the right subfigures are for Fe with $T_{\rm e} = 1.5$~keV. Two representative cases of the power-law index for the proton energy distribution are compared: $\Gamma = -1.5$, close to the standard index for the DSA (top), and $\Gamma = +1.0$, the index nearly maximizing the ionization rate for He-like ions (bottom). For all cases, the Bohr cross section is employed and the relative density of non-thermal protons to thermal electrons is set at $n_{\rm p}^{\rm nt} / n_{\rm e} = 1\%$. The lower panels of each subfigure show the ratio of ionic fractions, i.e., fractions with the proton ionization divided by those without it, in two different ranges.}
\label{fig:iconevo}
\end{figure*}

Figure~\ref{fig:iconevo} shows the time evolution of the ionic fractions for S and Fe using the Bohr cross section. Two representative cases are compared. One is with a power-law index of $\Gamma = -1.5$, close to the value expected in the standard diffusive shock acceleration  (DSA: \cite{Bell1978}) in the non-relativistic regime. The other is with a power-law index of $\Gamma = +1.0$, which nearly maximizes the ionization rates for He-like ions as shown in Figure~\ref{fig:rate_einddep}. The lower two panels of each subfigure in Figure~\ref{fig:iconevo} show ratios of ionic fractions with proton impact ionization to those without the proton effect. 
Overall, comparing the two different power-law index cases, the change in the CSD from a case without the proton ionization is larger with $\Gamma = -1.5$ for $\tau \lesssim 10^{11}$~cm$^{-3}$~s, while it is larger with $\Gamma = +1.0$ for $\tau \gtrsim 10^{11}$~cm$^{-3}$~s, for ions with significant fractions ($\gtrsim 0.01$). For a small $\tau$, ionization stage is lower than He-like, and thus the ionization is dominated by that at the 2s or higher shells, whose ionization rate is higher with a softer energy distribution. Once He-like ions dominate the CSD, the ionization rate is higher with a harder energy distribution because of the higher ionization threshold of the 1s shell. Since we assume a constant electron temperature fixed at $\approx T_{\rm e}^{\rm obs}$, the highest CSD for each parameter set is achieved when the ionization timescale reaches the CIE limit of $\tau \gtrsim 10^{12}$~cm$^{-3}$~s. 
With $n_{\rm p}^{\rm nt}/n_{\rm e} = 1$\%, the enhancement of ionic fraction of H-like S is by a factor of $\approx 7$ with $\Gamma = -1.5$, while it is $\approx 20$ with $\Gamma = +1.0$. The enhancement of H-like Fe is negligible with $\Gamma = -1.5$, but it is by a factor of $\approx 10$ with $\Gamma = +1.0$. With $\Gamma = +1.0$, the H-like to He-like ratio is $\approx 10$\% for S and $\approx 1$\% for Fe, which is smaller by a factor of a few compared to the observed ratios (Figure~\ref{fig:conrel}), indicating that the relative density of accelerated protons needs to be at least a few \%.

\begin{figure*}[htbp]
\begin{center}
\includegraphics[width=16cm]{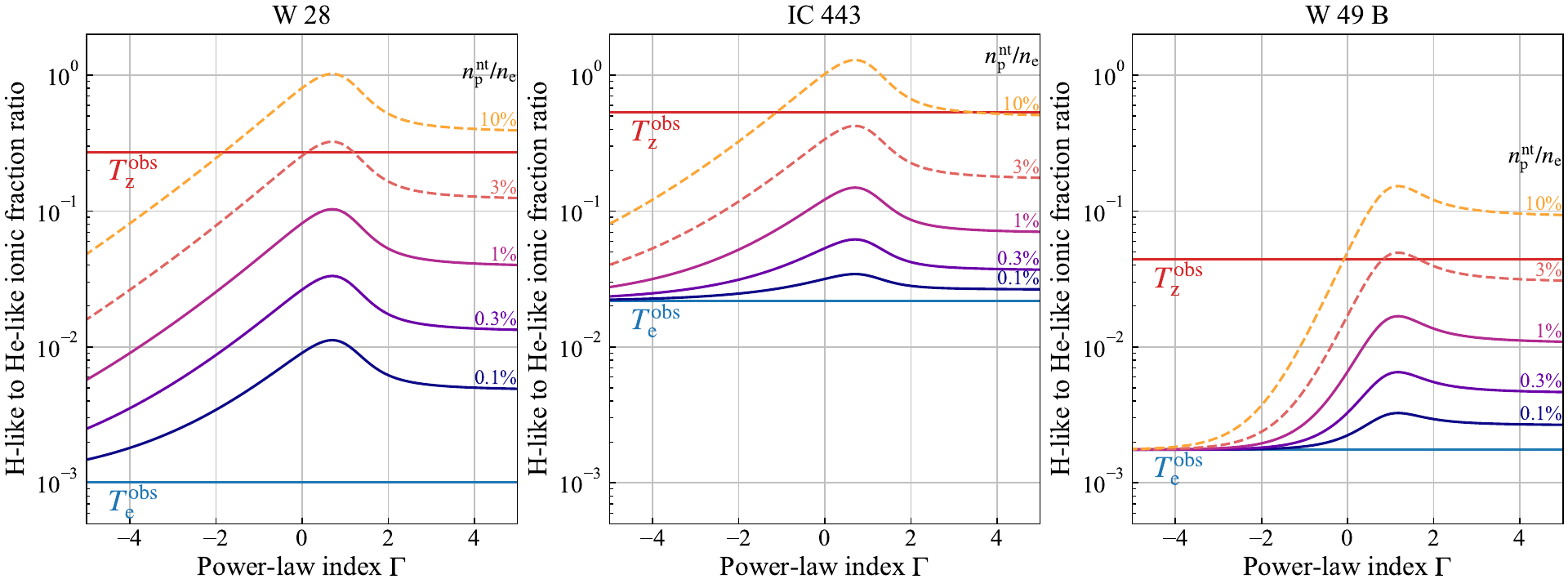}
\end{center}
\caption{The CSD at a large limit of the ionization timescale ($\tau$) with the proton impact ionization of the Bohr cross section, presented in the H-like to He-like ionic fraction ratio.
Assumed electron temperatures are of the observed values ($T_{\rm e}^{\rm obs}$) at the three recombining SNRs, W\,28 (left), IC\,443 (center), and W\,49\,B (right). Curves with different colors are of various densities of energetic protons relative to thermal electrons ($n_{\rm p}^{\rm nt}/n_{\rm e}$), each of which is given as a function of the power-law index of the proton energy distribution ($\Gamma$). The values at the observed ionization temperatures ($T_{\rm z}^{\rm obs}$) are given in the dark red horizontal lines.}
\label{fig:eqcsd}
\end{figure*}

Figure~\ref{fig:eqcsd} shows the maximum achievable CSD with the proton impact ionization at the large ionization timescale limit of $\tau \gg 10^{12}$~(cm$^{-3}$~s) for the three representative over-ionized SNRs. The Bohr cross section is used as it generally gives the highest proton ionization rates among the three models. In all cases, it is demonstrated that the relative density of accelerated protons to thermal electrons needs to be a few \% even with the optimal $\Gamma$ value of $\approx +1.0$ that maximizes the ionization rates for He-like ions.

\section{Discussion} \label{sec:disc}

\subsection{Requirement for proton ionization} \label{sec:disc1}

\subsubsection{Representative cases from observations}

\begin{figure}[htbp]
\begin{center}
\includegraphics[width=8cm]{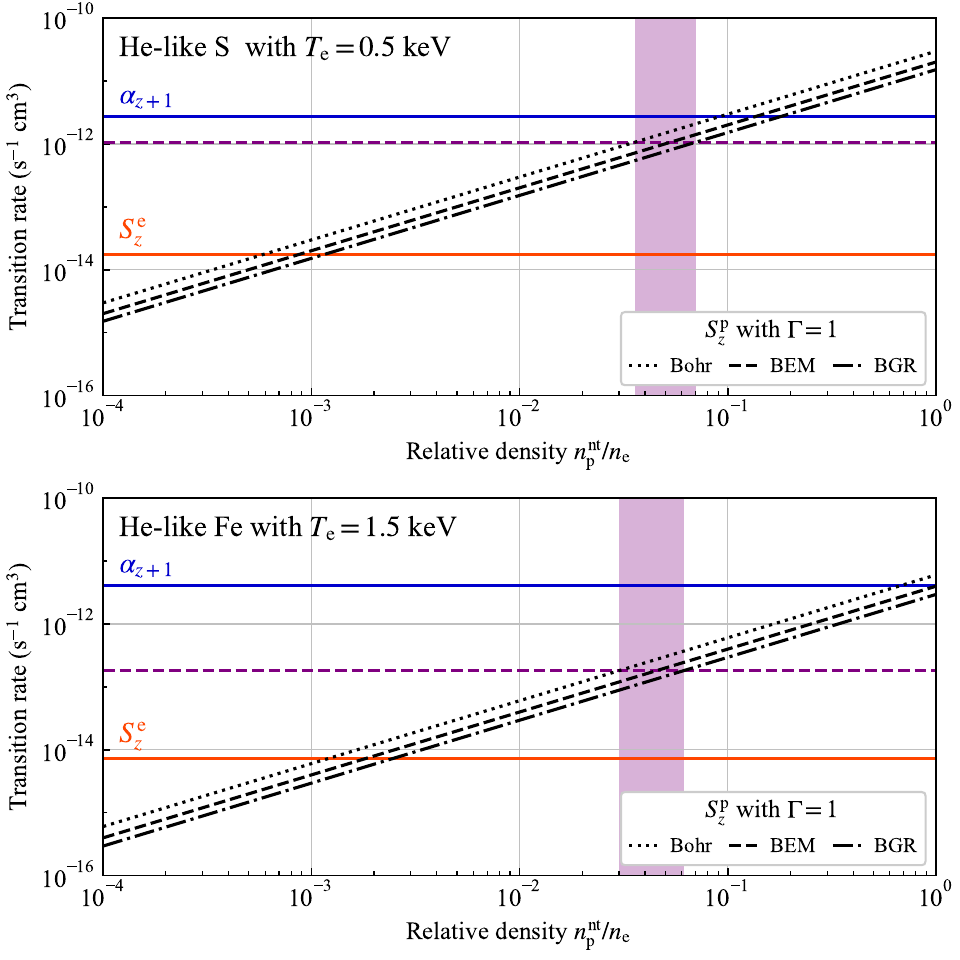}
\end{center}
\caption{The proton ionization rates for He-like S (top) and He-like Fe (bottom), each as a function of the relative density to thermal electrons. The power-law index $\Gamma = 1$ is assumed for the proton energy distribution. For comparison, the electron ionization rates of the same He-like ions with $T_{\rm e} = 0.5$~keV (top) and $T_{\rm e} = 1.5$~keV (bottom) are shown with the red horizontal lines, while the H-like to He-like recombination rates at the same temperatures are shown with the blue horizontal lines. Required ionization rates to explain the observed H-like to He-like fraction ratio (Figure~\ref{fig:conrel}) with the proton ionization at the CIE limit are shown with the dashed purple lines, while corresponding relative density ranges of accelerated protons to thermal electrons are indicated with purple areas.}
\label{fig:rate_relndep}
\end{figure}

With the power-law index fixed at $\Gamma = +1.0$ where the proton ionization rate is nearly maximized for He-like S and Fe (Figure~\ref{fig:rate_einddep}), the proton ionization rates are shown as a function of the relative density $n_{\rm p}^{\rm nt}/n_{\rm e}$ in Figure~\ref{fig:rate_relndep}. These are compared to the electron ionization and recombination rates with parameters representing IC\,443 and W\,28 ($T_{\rm e} = 0.5$~keV for He-like S) and W\,49\,B ($T_{\rm e} = 1.5$~keV for He-like Fe). An important outcome from this comparison is the CSD at the CIE limit, which gives the highest CSD in our time evolution model where the electron temperature is kept at a constant value. At CIE, $f_z \cdot S_z = f_{z+1} \cdot \alpha_{z+1}$, and thus the H-like to He-like ionic fraction ratio is given by

\begin{equation}
\frac{f_{\rm H\textrm{-}like}}{f_{\rm He\textrm{-}like}} \rightarrow \frac{S_{\rm He\textrm{-}like}}{\alpha_{\rm H\textrm{-}like}} = \frac{S_{\rm He\textrm{-}like}^{\rm p} + S_{\rm He\textrm{-}like}^{\rm e}}{\alpha_{\rm H\textrm{-}like}}. \label{eq:cielimit}
\end{equation}

\noindent
An IC\,443/W\,28-like case with $T_{\rm z}^{\rm obs} \approx 1.1$~keV corresponds to a H-like to He-like ratio for S of a few 10\% (Figure~\ref{fig:conrel} top), which, in combination with $\alpha_{\rm H-like}$, requires the relative density of $n_{\rm p}^{\rm nt}/n_{\rm e} \gtrsim$ a few \% (Figure~\ref{fig:rate_relndep} top). A W\,49\,B-like case with $T_{\rm z}^{\rm obs} \approx 2.7$~keV means a H-like to He-like ratio for Fe of a few \% (Figure~\ref{fig:conrel} bottom), again requiring the relative density of $n_{\rm p}^{\rm nt}/n_{\rm e} \gtrsim$ a few \% (Figure~\ref{fig:rate_relndep} bottom). These confirm the estimate in \S\ref{sec:ionfrac}.

\subsubsection{A more generalized requirement}

Here, we further attempt to derive a more generalized form of the minimum required density of accelerated protons for broad ranges of the electron temperature and elements. We concentrate on the electron temperature range where the He-like ions are dominant and examine the H-like to He-like ionic fraction ratio because all the robust detections of over-ionized plasma in SNRs have been made with He-like RRC. First, we evaluate the H-like to He-like ionic fraction ratio at a given electron temperature. Then, assuming the ionization to electron temperature ratio of 2 (a typical value for known recombining SNRs; Table~\ref{tab:obsts}), we evaluate the ionic fraction ratio at the corresponding ionization temperature ($T_{\rm z} = 2 T_{\rm e}$). The highest H-like to He-like ionic fraction ratio is realized at the CIE limit (\S\ref{sec:disc1}). From Eq.~\ref{eq:cielimit}, the condition that the proton ionization rate for He-like ions ($S_{\rm He\textrm{-}like}^{\rm p}$) should satisfy is given as follows:

\begin{equation}
\frac{S_{\rm He\textrm{-}like}^{\rm p} + S_{\rm He\textrm{-}like}^{\rm e}\left(T_{\rm e}\right)}{\alpha_{\rm H\textrm{-}like}\left(T_{\rm e}\right)} \ge \frac{S_{\rm He\textrm{-}like}^{\rm e}\left(T_{\rm z}\right)}{\alpha_{\rm H\textrm{-}like}\left(T_{\rm z}\right)}.
\end{equation}

The proton ionization rate $S_{\rm He\textrm{-}like}^{\rm p}$ is proportional to the relative density of accelerated protons to thermal electrons (Eq.~\ref{eq:sion_p}), and hence once the required proton ionization rate is determined, by comparing it to the modeled ionization rate (\S\ref{sec:model}), we can obtain the required relative density of accelerated protons. The modeled ionization rate varies depending on then choice on the cross section model and assumed power-law index $\Gamma$. We use the Bohr cross section as it gives the largest cross section. For the power-law index, we evaluate various values from $-5$ to $5$ with a step of 0.1 and choose the optimal one ($\Gamma_{\rm opt}$) at which the ionization cross section is maximized for each element.

\begin{figure}[htbp]
\begin{center}
\includegraphics[width=8cm]{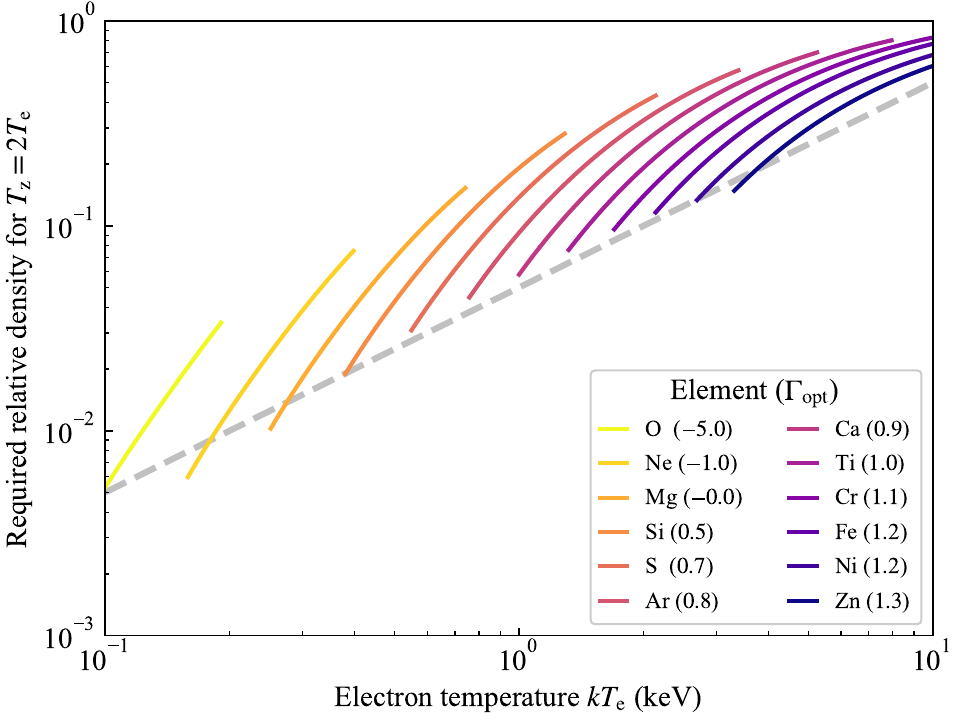}
\end{center}
\caption{Relative density of the non-thermal protons to thermal electrons required to raise H-like to He-like ionic fraction ratio to $T_{\rm z} = 2 T_{\rm e}$ using the Bohr proton ionization cross section. Curves are for different elements with the optimal power-law index ($\Gamma_{\rm opt}$) that maximizes the proton ionization rate for each element in a temperature range where He-like ionic fraction is greater than 1\%. A gray dashed line is an approximated requirement given as a function of the electron temperature, $5\,(kT_{\rm e}/{\rm 1~keV})$\,\%.}
\label{fig:minimum}
\end{figure}

The results is shown in Figure~\ref{fig:minimum}. Here, curves are limited to the electron temperature range in which the fraction of He-like ions is greater than 1\% for each element. By approximately connecting the minimal values of individual curves, we set a generalized requirement for the relative density of accelerated protons to thermal electrons to be 

\begin{equation}
n_{\rm p}^{\rm nt} / n_{\rm e} \gtrsim 5\,\left(\frac{kT_{\rm e}}{{\rm 1~keV}}\right)\,\%, \label{eq:reln_llim}
\end{equation}

\noindent
for an over-ionization at $T_{\rm z} = 2 T_{\rm e}$. We note that, this estimate is fairly robust because it depends only on the ionization and recombination rates of He-like and H-like systems. Unlike Li-like or lower ionization ions, these ions only have 1s electrons at the ground state, and thus they are affected by simple direct impact ionization only and free from complicated resonant or multiple-step processes such as excitation--auto-ionization and Auger effect after inner-shell ionization.

\subsection{The origin of recombining plasma} \label{sec:disc2}

\subsubsection{Proton ionization}

We have shown that, for the representative three cases of recombining SNRs, the observed ionization temperature can not be realized with the proton impact ionization unless the relative density of accelerated protons to thermal electrons exceeds a few \%, and in more general, a similar level of the relative density is required for broad ranges of electron temperature and elements (\S\ref{sec:disc1}). 

Considering that $n_{\rm e}/n_{\rm p} \approx 1.2$ for a solar abundance plasma or higher for heavy-element rich plasma, the relative density of accelerated protons to thermal electrons should be comparable to or smaller than the injection fraction of protons into acceleration, $\eta_{\rm inj}$,

\begin{equation}
\eta_{\rm inj} \equiv n_{\rm p}^{\rm nt} / n_{\rm p} \gtrsim n_{\rm p}^{\rm nt} / n_{\rm e}. \label{eq:reln_eta}
\end{equation}

\noindent
Here, the injection fraction is defined as the number density ratio of accelerated protons to total protons\footnote{Do not confuse the injection fraction and acceleration efficiency. The acceleration efficiency is rather defined as the relative energy flux, i.e., the ratio of the energy flux of accelerated particles to shock ram pressure. Depending on literatures, the symbol $\eta$ means either one or the other, e.g., \citet{Volk2003} vs. \citet{Kang2014}.}. Therefore, the feasibility to produce an over-ionized plasma by the proton impact ionization depends on how high the proton injection fraction is. 

The exact value of the proton injection fraction for SNR shocks is not known, but one may gauge its scale from a few parameters. The total non-thermal energy $E_{\rm tot}^{\rm nt}$ injected into accelerated particles is required to be $\sim 10$\% of the SN explosion energy $\sim 10^{51}$~erg to explain the observed Galactic cosmic ray flux by SNRs \citep{Blasi2013}. Thus, the number density of accelerated protons in ejecta can roughly be estimated by dividing it by the average energy per accelerated proton $\overline{E_{\rm p}^{\rm nt}}$ and the volume that confines accelerated protons: 

\begin{equation}
n_{\rm p}^{\rm nt} \sim 10^{-5} \left(\frac{E_{\rm tot}^{\rm nt}}{10^{50}~\rm erg}\right) \left(\frac{\overline{E_{\rm p}^{\rm nt}}}{0.1~\rm GeV}\right)^{-1} \left(\frac{R}{10~\rm pc}\right)^{-3}~~{\rm cm^{-3}},
\end{equation}

\noindent
where $R$ is the radius of the volume considered, whose typical value is taken to be $R \sim 10$~pc, the major axis of the shocked ejecta in the smallest over-ionized SNR,  W\,49\,B. The injection fraction is estimated to be $\eta_{\rm inj} \sim 10^{-5}$ for a typical thermal electron density of $n_{\rm e} \sim 1$~cm$^{-3}$. It may be much larger if  accelerated protons are confined in a small region; for instance, if we take $R = 1$~pc, then the $\eta_{\rm inj}$ may be as large as $10^{-2}$, which is comparable to the required relative density (\S\ref{sec:disc1}). However, such a small volume compared to observed extent of shocked ejecta implies that only a small fraction of thermal ions are affected by proton ionization, thus the over-ionization that dominates the observed spectra cannot be explained. 

Similar values of the injection fraction have been obtained in numerical simulations. As pointed out by \citet{Volk2003}, the injection fraction depends strongly on the shock obliquity. For a limit of a small shock inclination angle (parallel shock), a large injection fraction of $\eta_{\rm inj} \sim 10^{-2}$ is predicted by hybrid simulations (e.g., \cite{Scholer1992, Bennett1995}), which has been confirmed by measurements near the Earth's bow shock (\cite{Trattner1994}) and is consistent with expectations from analytical models \citep{MalkovVolk1995, MalkovVolk1998, Malkov1998}. The fraction is drastically suppressed at a larger shock inclination angle (quasi-perpendicular shock) because of a higher threshold energy above which protons are injected into DSA, which has also been demonstrated in general in kinetic simulations \citep{Caprioli2014, Caprioli2015}. For instance, although \citet{Caprioli2015} reported a large injection fraction of a few \% for the shock inclination of $\lesssim 45\degree$, it dropped rapidly and became smaller by one order of magnitude at the inclination of $50\degree$. Therefore, it is very unlikely that a large injection fraction of $\sim 1$\% is maintained on average in SNR shocks. 

These all indicate a conservative upper limit for the injection fraction averaged over an SNR to be

\begin{equation}
\eta_{\rm inj} \lesssim 10^{-2}. \label{eq:eta_ulim}
\end{equation}

From the above estimates (Eqs.~\ref{eq:reln_llim} and \ref{eq:eta_ulim}), it is clear that the required relative density of accelerated protons is too high to be realized in SNR shocks. Therefore, we conclude that the proton ionization scenario cannot explain the high over-ionization of $T_{\rm z}/T_{\rm e} \gtrsim 2$ found in Suzaku observations (e.g., \cite{Yamaguchi2009, Ozawa2009, Ohnishi2011, Sawada2012}). Note that actual conditions are even severer. For instance, we have assumed that the proton impact ionization lasts until the CSD reaches the CIE limit, but it is not the case if the injection fraction drops as an SNR evolves or the plasma density is not high enough to reach the CIE limit at the current age of an SNR. Also, we have assumed that the entire ejecta are affected by the additional ionization due to accelerated protons, but it is possible that only a fractional volume of the ejecta hosts such interaction. Finally, we have derived the requirement on the proton ionization such that it can explain the observed ionization temperature. However, spectral studies have repeatedly shown that the thermal relaxation of an over-ionized plasma toward a CIE state (the recombining phase in Figure~\ref{fig:schematic}) is not negligible, meaning that the ionization temperature achieved in the past would have been even higher than what we observe. The discrepancy between the required and available relative density of accelerated protons could easily builds up, say, by a factor of a few for each of these factors, making the proton ionization scenario even more unfeasible.

We have examined the ionization due to accelerated protons, but here, we briefly discuss another possibility of the proton ionization, which is contribution from thermal protons. As discussed above, in the case of accelerated protons, the small fraction of accelerated protons was a key limiting factor. Hence, one may naively expect that thermal protons, which have orders-of-magnitude higher relative densities, may have significant impact on CSD through their direct-impact collisional ionization of ions. In this case, however, the post-shock proton temperature needs to be unusually high. Numerical and analytical solutions of the Sedov model predict a speed of $v \sim 5  \times 10^3 \sqrt{E_{51} / M_{\rm ej}}$~km~s$^{-1}$ for the reverse shock at the rest frame of the pre-shock ejecta (e.g., \cite{Dwarkadas1998, Truelove1999}), where $E_{51}$ is the explosion energy in units of $10^{51}$~erg and $M_{\rm ej}$ is the ejecta mass in solar mass. Indeed, an application of such a model to the Galactic youngest SNR G\,1.9$+$0.3 \citep{Brose2019} yielded a reverse shock speed of 5650~km~s$^{-1}$ at the age of 140~yr. For protons, a corresponding post-shock temperature is $\sim 50$~keV with the Coulomb cooling ignored. At this temperature, the fraction of thermal protons above the energy at which the ionization cross section becomes large for He-like ions ($\gtrsim 1$~MeV for S and Fe: Figure~\ref{fig:xsec}) is negligibly small. A much higher temperature than this typical value would not be easily realized from the diversity of SN progenitors and explosions. For instance, there is a known positive correlation between these two quantities for Type II SNe (e.g., \cite{Hamuy2003, Pejcha2015}). In conclusion, as for the case of accelerated protons, the ionization by thermal protons are also unlikely to have significant impact on CSD of shock-heated ejecta.

\subsubsection{Electron cooling via conduction}

The other scenarios to explain the over-ionized plasma in SNRs rely on rapid electron cooling. One of the major scenarios is the conduction cooling by collisions with cold atomic or molecular clouds \citep{Kawasaki2002, Kawasaki2005}. Spatial correlation between a low $T_{\rm e}$ region and could interaction region has been found for a few over-ionized SNRs (e.g., \cite{Matsumura2017, Okon2018, Sano2021}). On the other hand, the systematic comparison of the thermal properties between over-ionized (recombining) and under-ionized (ionizing) SNRs has shown that the over-ionized SNRs systematically have higher ionization temperature rather than lower electron temperature compared to the under-ionized samples with similar dynamical ages \citep{Yamauchi2021}, which was the original motivation to introduce the proton ionization scenario. This systematic trend at least suggests that the conduction cooling alone cannot be the dominant process to explain the observed samples of over-ionized SNRs since it does not involve any process that raises the initial temperature before the cooling to make the observed ionization temperature higher than that in under-ionized SNRs.

An often overlooked fact is that, even if the conduction cooling by cold clouds is actually operating, it is not necessarily the origin of over-ionized plasma. If an over-ionized plasma is formed by another mechanism and then collides with nearby cold clouds, the same spatial correlation as reported can be reproduced. We note that, this interpretation of the spatial correlation is rather consistent with the systematic trend reported by \citet{Yamauchi2021}. We caution that, despite being an informative observational clue, a spatial correlation alone cannot provide firm evidence for the origin of over-ionized plasma unless there is a quantitative test that can distinguish whether the conduction cooling is the true origin or just a post process after the formation of the over-ionization state. 

\subsubsection{Electron cooling via rarefaction}

The other major cooling scenario is the rarefaction model, in which rapid adiabatic cooling is caused by a forward shock breaking out of dense circumstellar medium (CSM) left by a massive progenitor \citep{Itoh1989, Shimizu2012}. As for the conduction cooling scenario, a spatial correlation that supports this scenario has been found for W\,49\,B; regions with lower recombination timescales, implying lower electron densities, have lower electron temperatures \citep{Yamaguchi2018}. Although here we categorizes this as an electron cooling scenario, the consequences of the existence of dense CSM involve more than just the electron cooling. First, because of the dense environment outside the ejecta, the CSM mass shocked by the forward shock reaches more than 10\% of the ejecta mass at an age of only a few 10~yr, which is more than 10 times earlier than a remnant where no such dense CSM exists. As a result, the ejecta are shock-heated by a strong reverse shock and equipartition between thermal ions and electrons is quickly reached due to the high density of ejecta at this early stage. Once the forward shock breaks out of the CSM, then the rarefaction wave propagates into the CSM and ejecta, adiabatically cooling ions and electrons simultaneously but leaving the ionization temperature (or CSD) alone at a high value. In the rarefied hot ejecta, the electron density is no longer high enough to quickly attain the CIE condition, which makes the remnant observable as an over-ionized plasma for a long time (a characteristic recombination timescale of a few $\sim 10^5$~yr for a typical electron density of $n_{\rm e} \sim 0.1$~cm$^{-3}$ at an age of $\sim 10^3$~yr; \cite{Shimizu2012}). This scenario has a clear advantage over the conduction cooling scenario in the sense that it predicts a high initial temperature of ejecta and thus a high ionization temperature, which is consistent with the observed systematic trend where the difference between over-ionized and under-ionized SNRs is found in the ionization temperature rather than the electron temperature \citep{Yamauchi2021}. 

Based on the discussions above, in the present paper, we conclude that the rarefaction (adiabatic cooling) scenario would be the most plausible origin for the over-ionized (recombining) plasma in SNRs. One difficulty of this scenario would be the progenitor. The original proposal of this scenario assumed dense CSM from a red supergiant (RSG) star that explodes as a Type II core-collapse supernova, although some other types of a progenitor star could provide a similar environment (e.g., \cite{Moriya2012}). Identifications of explosion types and progenitor stars are not conclusive for all the over-ionized SNRs. For example, IC\,443 has a compact remnant (\cite{Swartz2015}) that indicates a Type II origin, but the progenitor of W\,49\,B is still in debate. Recent studies of W\,49\,B have reported abundance patterns of the ejecta that are more in line with a Type Ia origin (e.g., \cite{Zhou2018, Sato2024}), which seems to be inconsistent with a major premise of the rarefaction model. However, an exotic explosion scenario such as common envelope jets supernova explosion may simultaneously explain the Type-Ia-like abundance pattern and existence of dense CSM consisting of stellar winds from an RSG star \citep{Grichener2023}. 

\subsection{Prospects for XRISM}

\begin{figure}[htbp]
\begin{center}
\includegraphics[width=8cm]{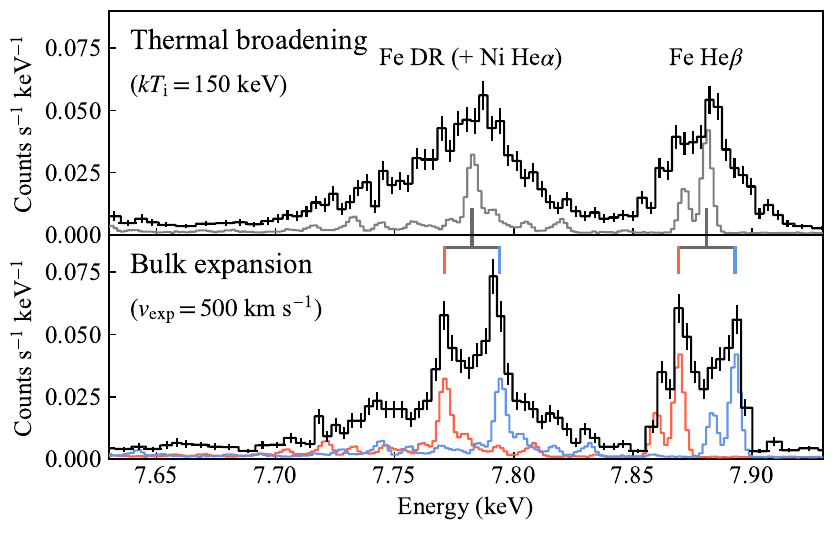}
\end{center}
\caption{Simulated spectra of the Fe He$\beta$ complex with XRISM/Resolve. Results with two different assumptions on the line broadening are shown: thermal motion dominant case (top) and bulk motion dominant case (bottom). The gray histogram in the top panel shows an unbroadened spectrum. The red and blue histograms in the bottom panel are the same but for redshifted and blueshifted velocity components along the line of sight that represent the expected velocity structure for the dynamical broadening dominant case. A total exposure time of 500~ks with the gate-valve closed configuration of the Resolve spectrometer is assumed.}
\label{fig:w49bheb}
\end{figure}

A more direct clue to distinguish different scenarios is expected to be obtained with high resolution X-ray spectroscopy from X-ray Imaging and Spectroscopy Mission (XRISM: \cite{Tashiro2022}). Figure~\ref{fig:w49bheb} shows simulated spectra of the Fe He$\beta$ complex of W\,49\,B with the Resolve microcalorimeter spectrometer \citep{Ishisaki2022} on XRISM, based on the spectral parameters from NuSTAR observations \citep{Yamaguchi2018} and surface brightness distribution from the Chandra 4.1--8.0~keV image\footnote{https://hea-www.cfa.harvard.edu/ChandraSNR/G043.3-00.2/}.
 With the energy resolution of 4.5~eV at full width half maximum at 6 keV, Resolve is capable of separate Fe\,\emissiontype{XXV} He$\beta$ (1s$^2$\,$^1$S$_0$--1s\,3p\,$^1$P$_1$ and 1s$^2$\,$^1$S$_0$--1s\,3p $^3$P$_1$) and Fe\,\emissiontype{XXIV} dielectronic recombination (DR) satellites (2p\,$^2$P$_{3/2}$--1s\,2p\,($^3$P)\,3p\,$^2$D$_{5/2}$ and 2p\,$^2$P$_{3/2}$--1s\,2p\,(3P)\,3p\,$^2$D$_{3/2}$). The Fe\,\emissiontype{XXV} He$\beta$ lines in particular have simple intrinsic line profiles, which allow us to investigate the broadening structure.  

Previous studies with X-ray CCDs have not detected any significant line broadening from W\,49\,B, and the statistical and systematic uncertainties would put a constraint at $\sigma_E \lesssim 10$~eV in the Fe-K band. If the conduction cooling scenario is the case, we expect a high ion temperature of $T_{\rm i} \gg T_{\rm e}$ because in this scenario no dense CSM is necessarily required, and thus, naively, an explosion in a uniform, low-density ISM can be assumed. In such an environment, the ion-electron equipartition timescale (a few $\sim 10^4$~yr for $n_{\rm e} \sim 1$~cm$^{-3}$) is longer than the age of the SNR (a few $10^3$~yr), inferring $T_{\rm i} \gg T_{\rm e}$ during the under-ionized phase before the cooling. Moreover, during the conduction cooling with cold clouds, it is expected that electrons are preferentially cooled due to a greater thermal conductivity than that for ions, which would increase the discrepancy between the ion and electron temperatures. Therefore, the line broadening would be predominantly by thermal broadening, as shown in the top panel of Figure~\ref{fig:w49bheb}, where we assume $T_{\rm i} = 150$~keV or $T_{\rm i} = 100\,T_{\rm e}$. 

A different line broadening is expected for the rarefaction (adiabatic cooling) scenario. In this scenario, an ion-electron equipartition is likely established at an early stage of the thermal evolution, and in a later stage, ions and electrons are cooled simultaneously. Thus, we expect $T_{\rm i} \approx T_{\rm e} = 1.5$~keV, which results in a thermal broadening of only $\sigma_E \approx 1$~eV for Fe K-shell lines. If there is any factor that could alter the line structure significantly, it would be dynamical effects, presumably reflecting a radial or axial expansion of the ejecta. To simulate a spectrum affected by dynamical effects, a simple model assuming a spherically symmetric expansion with a uniform expansion speed of $v_{\rm exp} = 500$~km~s$^{-1}$ is employed to get the line-of-sight velocity distribution as a function of the projected radius from an approximate center of the remnant. The resultant spectrum is shown in the bottom panel of Figure~\ref{fig:w49bheb}. Because of the centrally concentrated X-ray morphology, in this model, two major velocity components are expected, each represents redshifted or blueshifted one, with minor contribution from slightly different line-of-sight velocities that form non-Gaussian wings. Other than the line splitting and wings, the line profiles are narrow due to the ion temperature that is as low as the electron temperature, which would be a distinctive signature indicating the rarefaction due to dense CSM. Note that, the actual line structure depends on the dynamics of the ejecta, which are still unknown. For instance, if the bulk motion is more like bipolar flows (e.g., as one may naively expect from the bar-shaped morphology; \cite{Lopez2013}) rather than a radially expanding sphere or disk seen from an equatorial direction (e.g., as simulated by \cite{Shimizu2012}), then the line-of-sight velocity near the bright center region would be smaller, and hence the integrated spectrum may not show a clear bimodal structure as shown in Figure~\ref{fig:w49bheb}. Practically, spatially resolved spectroscopy would be needed to distinguish the dominant cause of the broadening and eventually to constrain the ion temperature.

\section{Summary} \label{sec:summary}

The impact of including additional ionization of SNR ejecta due to the low-energy tail of the accelerated protons was studied. We first established simple models for thermal evolution and CSD (\S\ref{sec:model}). Then we formulated the proton direct impact ionization cross sections using relevant scalings from the electron direct impact ionization cross sections (\S\ref{sec:xsec}). Using these, we evaluated the ionization rates and obtained the time evolution of CSD for various elements (\S\ref{sec:result}). We found that, for representative examples of over-ionized SNRs, accelerated protons need to be as abundant as a few \% of thermal electrons to explain the observed discrepancy of the ionization temperature from electron temperature. The discussions first focused on the impact of the proton ionization on CSD for representative cases of known over-ionized SNRs (\S\ref{sec:disc1}). We then derived a more generalized requirement on the minimum relative density of accelerated protons to reproduce observed high ionization temperatures. In the later part, we discussed the feasibilities of the major scenarios for producing over-ionized plasma in SNRs, including the proton ionization scenario (\S\ref{sec:disc2}). We found that the required relative density of accelerated protons is too high to be realized by an expected injection fraction at SNR shocks, and concluded that the proton ionization scenario to be unfeasible. We further discussed electron cooling scenarios. By taking previous observational results into account, we concluded that the rarefaction model would be the most plausible scenario. Nonetheless, a more direct and conclusive observational clue to distinguish proposed scenarios is greatly anticipated, which will hopefully be provided from high-resolution spectroscopy with XRISM.

\begin{ack}
We acknowledge Prof. Jelle Kaastra for early discussion on ionization rate calculations, Prof. Hiroya Yamaguchi for providing the NuSTAR spectral parameters of W\,49\,B, and Prof. Shunji Kitamoto for improving the draft. This research was partially supported by the RIKEN Pioneering Project Evolution of Matter in the Universe (r-EMU) and Rikkyo University Special Fund for Research (Rikkyo SFR) (M.\,S.), and JSPS KAKENHI Grant Numbers, JP22H01251, JP23H01211, JP23K22522, and JP23H04899 (R.\,Y.).

\end{ack}

\bibliographystyle{aa} 
\bibliography{ms} 

\end{document}